\documentclass{aa}  
\usepackage{graphicx}
\usepackage{txfonts}
\usepackage{textcomp}

\begin{document}

\title{The tiny globulettes in the Carina Nebula
          \thanks{Based on observations collected with the NASA/ESA Hubble Space Telescope, obtained at the Space Telescope Science Institute. 
} }


   \author{T. Grenman\inst{1}
           \and
	   G. F. Gahm\inst{2}
	   }


   \institute{Applied Physics, Department of Engineering Sciences \& Mathematics, Lule{\aa} University of Technology, SE-971 87 Lule\aa, Sweden\\
              email: \mbox{tiia.grenman@gmail.com}
              \and
	     Stockholm Observatory, AlbaNova University Centre, Stockholm University,
	      SE-106 91 Stockholm, Sweden   }

   \date{}

   

 \abstract
   {Small molecular cloudlets are abundant in many \ion{H}{ii} regions surrounding newborn stellar clusters. In optical images these so-called globulettes appear as dark silhouettes against the bright nebular background.} 
   {We aim to make an inventory of the population of globulettes in the Carina Nebula complex, and to derive sizes and masses for comparisons with similar objects found in other \ion{H}{ii} regions. }
   {The globulettes were identified from H$\alpha$ images collected at the Hubble Space Telescope.}
   {We have located close to 300 globulettes in the Carina complex, more than in any other region surveyed so far. The objects appear as well-confined dense clumps and, as a rule, lack thinner envelopes and tails. Objects with bright rims are in the minority, but more abundant than in other regions surveyed. Some globulettes are slightly elongated with their major axes oriented in the direction of young clusters in the complex. Many objects are quite isolated and reside at projected distances $>$~1.5 pc from other molecular structures in the neighbourhood. No globulette coincides in position with recognized pre-main-sequence objects in the area. The objects are systematically much smaller, less massive, and much denser than those surveyed in other \ion{H}{ii} regions. Practically all globulettes are of planetary mass, and most have masses less than one Jupiter mass. The average number densities exceed 10$^{5}$~cm$^{-3}$ in several objects. We have found a statistical relation between density and radius (mass) in the sense that the smallest objects are also the densest.}
  {The population of small globulettes in Carina appears to represent a more advanced evolutionary state than those investigated in other \ion{H}{ii} regions. The objects are subject to erosion in the intense radiation field, which would lead to a removal of any thinner envelope and an unveiling of the core, which becomes more compact with time. We discuss the possibility that the core may become gravitationally unstable, in which case free-floating planetary mass objects can form. }


 \maketitle
%

\section{Introduction}
\label{sec:intro}

Optical images of galactic \ion{H}{ii} regions show a mix of bright and dark nebulosity.  Foreground cold dust clouds obscure the bright background of warm and ionized gas. The warm plasma in these nebul\ae\ accelerates outwards through the interaction with radiation and winds from hot and massive stars. The molecular gas is swept up and forms expanding shells, which are sculpted into complex filamentary formations, like elephant trunks,  pillars that point at O stars in the nebula. Blocks of cold gas can detach from the shells and trunks and may fragment into smaller clouds that appear as dark patches on optical images of these nebul\ae\ as noted long ago by Bok \& Reilly (\cite{bok47}) and Thackeray (\cite{tha50}). The clumps may be round or shaped like tear drops, some with bright rims facing the central cluster (Herbig \cite{her74}).

A number of such \ion{H}{ii} regions have been subject to more detailed studies, and it has been found that many regions contain distinct, but very small clumps, extending  over less than one to a few arcseconds. Several studies have focused on the so-called proplyds, which are photoevaporating discs surrounding very young stars (e.g. O'Dell et al. \cite{ode93}; O'Dell \& Wen \cite{ode94}; McCaughrean \& O'Dell \cite{mcc96}; Bally et al. \cite{bal00}; Smith et al. \cite{smi03}). In these studies small cloudlets without any obvious central stellar objects were also found, as also recognized by Hester et al. (\cite{hes96}) and Reipurth et al. (\cite{rei97, rei03}) from Hubble Space Telescope images of nebular regions. 

More systematic studies of such star-less cloudlets followed, and from the surveys of more than 20 \ion{H}{ii} regions by De Marco et al. (\cite{mar06}), Grenman (\cite{gre06}), and Gahm et al. (\cite{gah07}; hereafter Paper~1) it can be concluded that most of the objects have radii $<$10~kAU with size distributions that peak at $\sim$~2.5~kAU. In Paper~1 masses were derived from extinction measures indicating that most objects have masses $<$~13~M$_{J}$ (Jupiter masses), which currently is taken to be the domain of planetary-mass objects. This class of tiny clouds in \ion{H}{ii} regions were called {\it globulettes} in Paper~1 to distinguish them from proplyds and the much larger globules spread throughout interstellar space. We define globulettes as cloudlets with round or slightly elongated shapes with or without bright rims and/or tails. 

Some globulettes are connected by thin filaments to larger molecular blocks and it is then natural to assume that isolated globulettes once detached from shells and trunks. They may also survive in this harsh environment for long times, as concluded in Paper~1. Follow-up 3D numerical simulations in Kuutmann (\cite{kuu07}) predict lifetimes of $\sim$~$10^{4}$ years, increasing with mass. Owing to the outer pressure exerted on the globulettes from surrounding warm gas, and the penetrating shock generated by photoionization, it was found that many globulettes may even collapse to form brown dwarfs or planetary-mass objects before evaporation has proceeded very far. The objects are protected against rapid photoevaporation by a screen of expanding ionized gas (e.g. Dyson \cite{dys68}; Kahn \cite{kah69}; Tenorio-Tagle \cite{ten77}). Consequently, the objects are expected to develop bright rims on the side facing the cluster because of the interaction with stellar light. In addition, the models predict that dusty tails emerge from the cloud cores. It is therefore puzzling that most globulettes lack any trace of bright rims in H$\alpha$, and that most are round, or only slightly elongated, without any trace of tails. 

In a recent study by Gahm et al. (\cite{gah13}; hereafter Paper~2), based on NIR imaging and radio molecular line observations of globulettes in the Rosette Nebula, it was found that the objects contain dense cores, which strengthens the suggestion that many objects might collapse to form planetary-mass objects or brown dwarfs that are accelerated outwards from the nebular complex. The whole system of globulettes and trunks expands outwards from the central cluster with velocities of about 22 km s$^{-1}$. In the case where more compact objects are formed inside some globulettes, they will escape and become free-floating objects in the galaxy. In both the optical and radio/NIR surveys (Papers 1 and 2) it was concluded that the density is relatively high even close to the surface layers, which could explain why the objects lack extensive bright rims in H$\alpha$. Some of the optically completely dark objects were discovered to have thin rims manifested in P$\beta$ and H$_{2}$ emission. In a follow-up study of the NIR images, M\"akel\"a et al. (\cite{mak14}) found that some smaller globulettes are also crowned by thin bright rims that are not seen in H$\alpha$. 

The present study is an inventory of globulettes in the Carina Nebula (NGC 3372) based on images taken from the {\it Hubble Space Telescope} (HST) through a narrow-band H$\alpha$ filter. Basic parameters, like size and mass, are derived and we compare the results to surveys of similar objects in other nebulae.

The Carina complex, with its extended network of bright and dark nebulosity, spans over several degrees in the sky and is one of the most prominent sites of star formation in the galaxy. More than 60 O-type stars and several young clusters (Tr 14, 15, and 16; Collinder 228 and 232; and Bochum 10 and 11) are located in the region, and more than a thousand pre-main sequence stars have been identified from optical, infrared, and X-ray surveys (e.g. Tapia et al. \cite{tap03}; Ascenso et al. \cite{asc07}; Sanchawala et al. \cite{san07a}, \cite{san07b}; Smith et al. \cite{smi10a}, \cite{smi10b}; Povich et al. \cite{pov11}; Gaczkowski et al. \cite{gac13}). The global properties of the nebular material was discussed in Smith et al. (\cite{smi00}), Smith \& Brooks (\cite{smi07}), and references to studies based on observations of selected areas can be found in the comprehensive review by Smith (\cite{smi08}). Additional surveys from the submm range (Preibisch et al. \cite{pre11}; Pekruhl et al. \cite{pek13}) and the far IR (Preibisch et al. \cite{pre12}; Roccatagliata et al. \cite{roc13}) have been made more recently. The Carina Nebula, in all its glory, is presented in multicolour mosaics found at the Hubble Space Heritage webpage.

A number of small obscuring structures in the Carina Nebula were noted by Smith et al. (\cite{smi03}) from HST images, and were regarded as possible proplyds. However, the objects studied were found to be larger than the standard cases in the Orion Nebula (Bally et al. \cite{bal00}). More objects of this nature were recognized by Smith et al. (\cite{smi04}) who stated that their nature remains ambiguous: "analogues of Orion's proplyds, starless cometary clouds, or something in between?" Ascenso et al. (\cite{asc07}), however, concluded from near-infrared imaging that these candidates do not harbour any stars. Most of these objects are globulettes by our definition and are thereby included in our list of nearly 300 globulettes. Thus the Carina complex is the richest known with regard to total number of globulettes. A number of Herbig-Haro jets emanating from embedded young stars in the region were found by Smith et al. (\cite{smi10a}). Most of these are related to trunks or larger fragments. However, HH 1006 is related to an isolated cloud with an embedded jet-driving source (Sahai et al. \cite{sah12}; Reiter \& Smith \cite{reit13}). Tentative jet-signatures were also found for a few much smaller isolated clouds like HH 1011 and HHc-1.

The distance to the Carina complex has been estimated in several investigations with rather different results. A distance of 2.3 kpc has been adopted as a kind of standard (Smith \cite{smi08}). Recently, Hur et al. (\cite{hur12}) concluded that the main stellar clusters Tr 14 and 16 are located at a distance of 2.9 kpc. We have adopted this value in the present investigation, but will discuss the implications if the complex is closer. 

The paper is organized as follows. We present the fields we have searched, the objects identified, and their measured properties in Section~\ref{sec:obs}. The results are analysed in Section~\ref{sec:results} and discussed further in Section~\ref{sec:disc}. We end with a summary in Section~\ref{sec:conclude}. 

\begin{figure}[t] 
\centering
\resizebox{9cm}{!}{\includegraphics[angle=00]{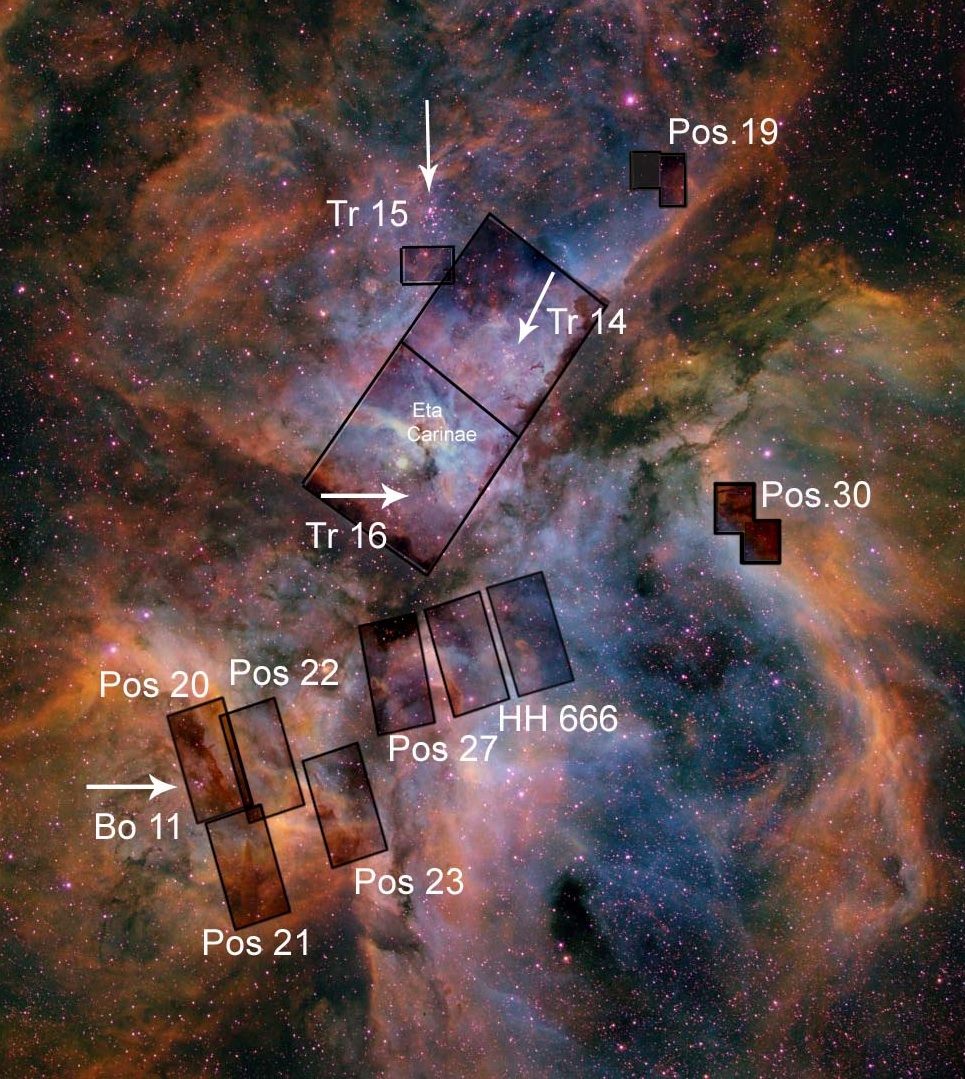}}
\caption{Image of the central region of the Carina Nebula, where the HST fields containing globulettes are marked. The locations of the star $\eta$~Carin{\ae} and four stellar clusters are marked. North is up and east to the left. The image spans 1.\degr3 x 1.\degr5  (credit: Nathan Smith, Univ. of Minnesota, NOAO, AURA, NSF).} 
\label{map}
\end{figure} 

\begin{table}
      \caption[]{HST archive data used.}
         \label{HST}
     $$
         \begin{array}{*{4}{p{0.1\textwidth}}}
            \hline
            \noalign{\smallskip}
            \ Field/Target & R.A. (J2000.0) \hfill{} &  Dec. (J2000.0)& Images \\
            \noalign{\smallskip}
            \hline
            \noalign{\smallskip}
1 / Pos 30 &   10:41:27	& 	-59:47:42  &  J9dk09010	 \\
2 / Pos 30 &   10:41:38	&   -59:46:17   &  J9dk09020    \\
3 / Pos 30 &   10:41:40	&   -59:44:41   &   J9dka9010   \\
4 / Pos 19 & 10:42:23   &-59:20:59     &   J9dk12010   \\
5 / Pos 19 & 10:42:48	&-59:19:44    &  J9dk32010    \\
6 / Tr 14   &  10:43:07	&   -59:29:34   &  J900c1010     \\  
7 / Tr 14 &   10:43:23	& 	-59:32:06  &  J900b1010 	 \\
8 / Tr 14&   10:43:24	 	& -59:27:55  & J900c2010     \\ 
9 / Tr 14 &   10:43:39	& 	-59:34:37   &  J900a1010    \\ 
10 / Tr14 &   10:43:41	& 	-59:30:17  &    J900b2010   \\ 
11 / Tr 14 &   10:43:47	& 	-59:35:53  &    J90001020   \\ 
12 / Tr 14 &   10:43:55	& 	-59:37:09   &   J90001010    \\ 
13 / HH 666&   10:43:58	& 	-59:54:39 &  J900a9010     \\
14 / Tr 14 &   10:43:59	& 	-59:32:38   &    J900a2010   \\ 
15 / Tr 14 &   10:44:00	& 	-59:28:05  &    J900b3010 	 \\
16 / HH 666 &   10:44:01	& 	-59:58:42  &    J900b9010   \\ 
17 / Tr 16 &   10:44:05	& 	-59:40:16  &    J900c5010   \\ 
18 / Tr 14 &   10:44:07	& 	-59:33:53 &    J90002020   \\ 
19 / Tr 14 &   10:44:15	& 	-59:35:09  &    J90002010   \\ 
20 / Tr 14 &   10:44:17	& 	-59:30:27  &    J900a3010   \\ 
21 / Tr 14 &   10:44:19	& 	-59:25:54  &    J900b4010   \\ 
22 / Tr 16 &   10:44:20	& 	-59:42:48  &    J900b5010   \\ 
23 / Tr 16 &   10:44:22	& 	-59:38:34 &    J900c6010   \\ 
24 / Tr 14 &   10:44:35	& 	-59:33:09 &    J90003010   \\ 
25 / Tr 16 &   10:44:36	& 	-59:45:19  &    J900a5010   \\ 
26 / Pos 27 &   10:44:40	& 	-59:59:46  &    J9dk07010   \\ 
27 / Pos 27&   10:44:43	& 	-59:56:34  &   J9dk27010    \\
28 / Tr 16  &    10:44:44 	& 	-59:46:35  &   J90005020    \\
29 / Tr 16 &   10:44:49	& 	-59:37:35  &    J900b7020   \\
30 / Tr 16 &   10:44:52	& 	-59:47:51 &    J90005010   \\ 
31 / Tr 14 &   10:44:54	& 	-59:31:08  &    J90004010   \\ 
32 / Tr 15 &   10:44:58	& 	-59:26:50  &    J9dka0010   \\ 
33 / Tr 16&   10:44:58	& 	-59:38:47  &    J900b7010   \\ 
34 / Tr 16 &   10:45:12	& 	-59:45:51 &    J90006010   \\
35 / Tr 16  &   10:45:17	& 	-59:36:38 &    J900b8010   \\
36 / Tr 15 &   10:45:23	& 	-59:26:59  &    J9dk10010   \\ 
37 / Tr 16 &   10:45:44	& 	-59:40:34  &    J90008020   \\ 
38 / Pos 23&   10:45:53   &  -60:08:16 	&   j90010010    \\ 
39 / Pos 23 &   10:45:56	& 	-60:06:42  &    J90010020   \\ 
40 / Pos 22&   10:46:32	& 	-60:05:14  &   J9dk23010    \\
41 / Pos 21&   10:46:47	& 	-60:09:29  &   J9dk22010    \\
42 / Pos 20&   10:46:58	& 	-60:06:26  &    J9dk01010 	 \\
43 / Pos 20&   10:47:01	& 	-60:03:14  &    J9dk21010   \\ 
            \noalign{\smallskip}
            \hline
         \end{array}
    $$
  \end{table}

\section{Objects and measurements}
\label{sec:obs}

\subsection{HST fields}
\label{sec:fields}

\begin{figure}[t]
\centering
\includegraphics[angle=00, width=9cm]{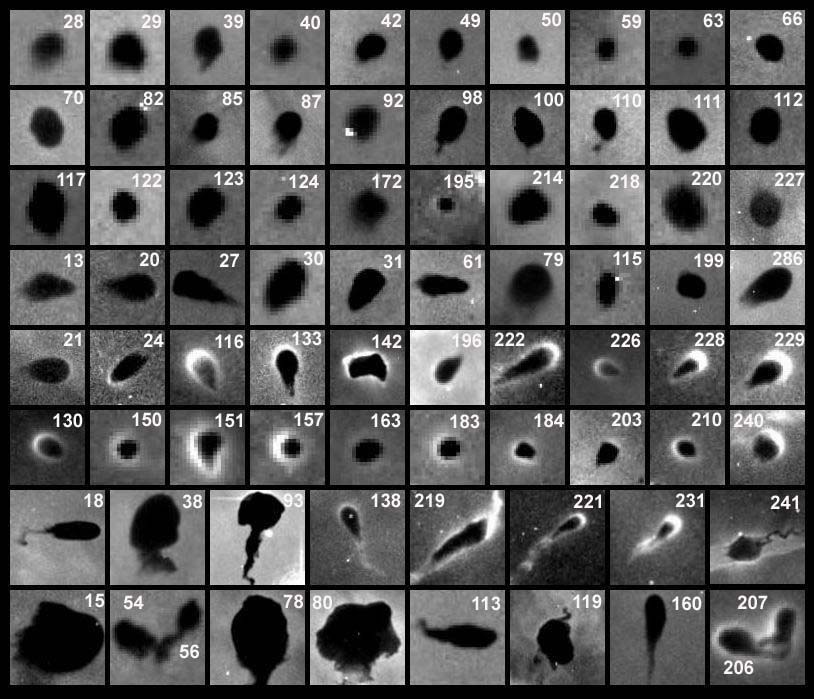}
\caption{Examples of globulettes found in the Carina Nebula numbered according to Tables A1-A5 in Appendix A. The most typical cases are the dark globulettes shown in the upper four rows followed by objects with bright halos. The last two rows show examples of elongated objects with tails, some with bright rims. We note the scales are different from panel to panel (the dimensions in arcsec are given for each object in the tables). }
\label{mosaik}
\end{figure}

The optical images of the Carina complex were downloaded from the HST archive, cycle 13 and 14 programs GO-10241 and 10475 (principal investigator N. Smith) based on observations with the ACS/WFI camera, which contains two CCDs of 2048~$\times$~4096 pixels glued together with a small gap in between. The pixel size corresponds to $\approx$~0.05 arcsec pixel$^{-1}$, and the field of view is $202 \times 202$~arcsec. All images selected were exposed for 1000~s through the narrow-band filter F658N, covering the nebular emission lines of H$\alpha$ and [\ion{N}{ii}]. 

Most of the HST fields contains globulettes, and only these are listed in Table~\ref{HST} with a running field number and references to target and image designations according to the HST archive.  Figure~\ref{map} shows how the areas covered by HST are distributed over the region (see also Smith et al. \cite{smi10a}). Two regions on opposite sides of a large V-shaped dark cloud are rather well covered by HST. The total area covered is $\sim$~700~arcmin$^2$, which is larger than covered by all HST-based surveys of \ion{H}{ii} regions together, but much smaller than the area covered of the Rosette Nebula in Paper~1. This ground-based survey was limited to objects with radii $\geq$~0.8 arcsec, however.  

The globulettes are easily recognized as dark patches against the bright background. Most are roundish without any bright rims or halos, similar to previous surveys. A number of the elongated objects with tear-drop forms are crowned with bright rims. The Carina complex is also rich in dark irregular blocks and fragments of all sizes, some of which are very elongated and shaped like worms or long, narrow cylinders, and some show very irregular shapes. These objects, which as a rule are much larger than typical globulettes, were not included in our list of globulettes, but in  Sect.~\ref{sec:pec} we highlight some smaller cloudlets with peculiar shapes.

The Carina Nebula contains a large number of very small globulettes, down to the limit of resolution of HST, but we do not consider objects with dimensions $\le$ 3 pixels across. Some regions contain quite isolated globulettes that are located far away from any larger molecular block, while in others there are clusters of globulettes. Examples of different types of globulettes are shown in  Fig.~\ref{mosaik}, where the first four rows show round and dark globulettes, which are most abundant. Round objects with bright halos are found in the fifth row followed by objects that are more elongated or have developed pronounced tails, with or without bright rims.

\subsection{Measurements}
\label{sec:data}

\begin{table*}
\centering
\caption{List of globulettes. The symbols are defined in Sect.~\ref{sec:data}. The complete list of the 288 globulettes measured is presented in Appendix A.}
\begin{tabular} {lccccccccccl}
 \hline
     \noalign{\smallskip}
CN & Field & x & y & R.A. & Dec. & $\alpha$ & $\beta$ & P.A. & $\bar r $ & Mass & Remarks\\
 &&&& (J2000.0) & (J2000.0) & (arcsec) &  (arcsec) & (degr.)  & (kAU) & (M$_{J}$) & \\

\noalign{\smallskip}   
\hline
\\
    \noalign{\smallskip}

1   &  F1  & 360   & 4020   & 10:41:13.3 & -59:49:00  & 0.46 & 0.50    &    &   1.39    & 2.4    &    \\                                
2   &  F1  & 660	 & 1084   & 10:41:18.9 & -59:46:38  & 0.30 & 0.36    &    &   0.96    & 1.1    &     \\
3   &  F1  & 1792  & 3251   & 10:41:23.6 & -59:48:35   & 0.38 & 0.40   &    &   1.13    & 1.3    &    \\
4   &  F1  & 1654  & 378    & 10:41:26.2 & -59:46:13   & 0.18 & 0.19   &    &   0.54    & 0.4    &    \\
5   &  F1  & 1819  & 395    & 10:41:27.2 & -59:46:15   & 0.19 & 0.27   &    &   0.67    & 0.6    &    \\
6   &  F1  & 2029  & 221    & 10:41:28.8 & -59:46:09   & 0.26 & 0.33   & 23   &   0.86    & 1.2    &    T  \\
7   &  F2  & 670	 & 1907   & 10:41:28.9 & -59:45:53   & 0.23 & 0.25   & -43&   0.69    &\it 0.6 &    BR,T \\
8   &  F1  & 2391  & 1935   & 10:41:29.1 & -59:47:36   & 0.20 & 0.22   &    &   0.61    & 0.5    &     \\
9   &  F1  & 2156  & 597    & 10:41:29.2 & -59:46:29   & 0.29 & 0.33   &    &   0.90    & 1.0    &    \\
10  &  F1  & 2752  & 1904   & 10:41:31.5 & -59:47:38   & 0.54 & 1.46   & 8 &   2.90    & \it 8.8    &  BR,EL,T\\
11  &  F1  & 2604  & 373	 & 10:41:32.4 & -59:46:22   & 0.31 & 0.51   & -38 &   1.19    & 2.0    &   EL  \\
\\
continued in Appendix A\\
\\
\hline

\end{tabular}
\label{glob} 
\end{table*} 

Central positions were measured in terms of x and y coordinates and R.A. and Dec according to available HST readouts. The globulettes, designated CN (as Carina Nebula plus number), are listed in order of increasing R.A. in Table 2 showing only the first entries. The complete table is found in Tables A.1-A.5 in Appendix A. Finding charts for all fields containing globulettes are found in Figs. B.1-B.6 in Appendix B. In these charts we have also marked some objects that we do not consider to be regular globulettes, like some clumps with peculiar shapes (see Sect.~\ref{sec:pec}). Some larger fragments are marked as \emph{Frag} and these features will be commented on in Sect.~\ref{sec:disc}. Most globulettes have circular or slightly elliptic shapes. The semi-major and semi-minor axes are given in arcseconds in Cols. 7 and 8. These quantites are defined from an outer contour where the intensity level has dropped to 95 \% of the interpolated background nebular intensity. Outside this contour, the level of noise starts to affect the definition of the boundary, but as a rule very little matter resides in the outskirts. Column 9 gives the position angle of elongated objects, for which the ratio of semi-major and semi-minor axes is $>$ 1.5.  

We derive the physical dimensions of the objects assuming a distance of 2.9 kpc (see Sect.~\ref{sec:intro}) and define a characteristic radius, $\bar r$, as the mean of the semi-major and semi-minor axes expressed in kAU (Column 10). For the determination of mass we strictly follow the procedure as described in detail in Paper~1 and Grenman (\cite{gre06}). In short, we measure the residual intensity for each pixel within a globulette relative to the interpolated bright background. This value relates to extinction due to dust at $\lambda$~6563~{\AA} ($A_{\alpha}$). Two extreme cases are considered: there is no foreground emission at all, or practically all the residual intensity in the darkest areas of each object is caused by foreground emission. We assume a standard interstellar reddening law (Savage \& Mathis \cite{sav79}) to compute the visual extinction, $A_{V}$ = 1.20 $A_{\alpha}$, and the column densities of molecular hydrogen, $N(H_{2})$ = 9.4~10$^{20} A_V$, according to the relations in Bohlin et al. (\cite{boh78}) for each pixel assuming a standard mass ratio of gas to dust of 100, and that all hydrogen is in molecular form. The total column density is derived assuming a cosmic chemical composition. Finally, we sum over all pixels inside the contour defined above to obtain the total mass, and we select the mean of the two extreme cases defined above as a measure of the mass of each object. Column 11 gives the so derived mean mass of each globulette. The maximum and minimum masses rarely differ from the mean by more than a factor of two. 

In the last column remarks about individual objects are found. Elongated globulettes are marked as $EL$. Some objects have developed tails or tear-drop forms and are marked $T$. Objects with pronounced bright rims are marked $BR$, and those with bright halos as $BH$. The derived masses for the $BH$ objects are lower limits, and their masses are set in italics in Column 11. Symbol $C$ indicates that the object is connected by a dark, thin filament to a larger structure, like a nearby trunk, or to another globulette (with number marked). Objects noted in Smith et al. (\cite{smi03}) are marked $S$ in Column 12 followed by the symbol they used, and two HH candidates recognized in Smith et al. (\cite{smi10a}) are also noted.

The derived masses are subject to other uncertainties as well. For instance, uniform density has been assumed, which is consistent (to a first approximation) with column densities derived as a function of radial position (see Paper 1). However, the objects may have developed dense cores that escape detection. Another concern is the use of a normal extinction law since larger-than-normal ratios of $R$ have been found in certain areas (e.g. Th\'e et al. \cite{the80}; Smith \cite{smi02}; Tapia et al. \cite{tap03}; Hur et al. \cite{hur12}). Since the globulettes may condense from larger clouds, they may contain larger dust grains than assumed for a normal extinction law. Finally, nebular H$\alpha$ photons entering a globulette may scatter into the line of sight to the observer (e.g. Mattila et al. \cite{mat07}). This effect would lead to an underestimation of mass. The effect is expected to be small, but cannot be evaluated further until more precise information exists on locations within the nebula and local radiation fields.

\begin{figure*}[t] 
\centering
\resizebox{6cm}{!}{\includegraphics[angle=00]{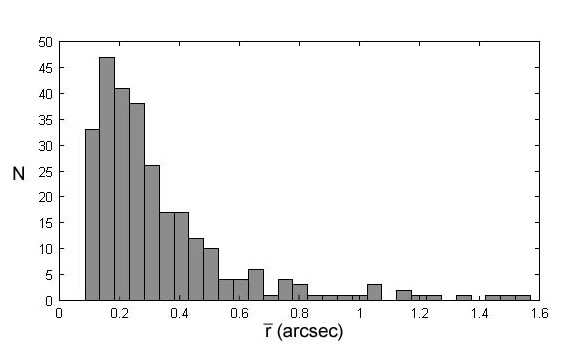}}
\resizebox{6cm}{!}{\includegraphics[angle=00]{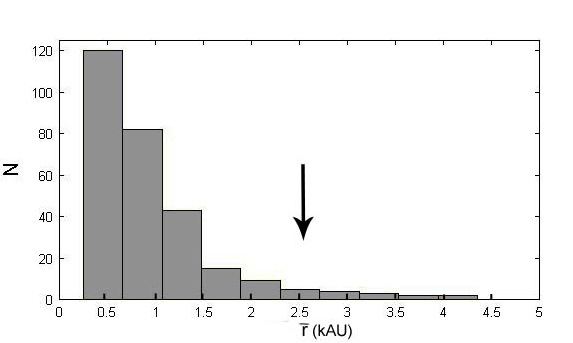}}
\resizebox{6cm}{!}{\includegraphics[angle=00]{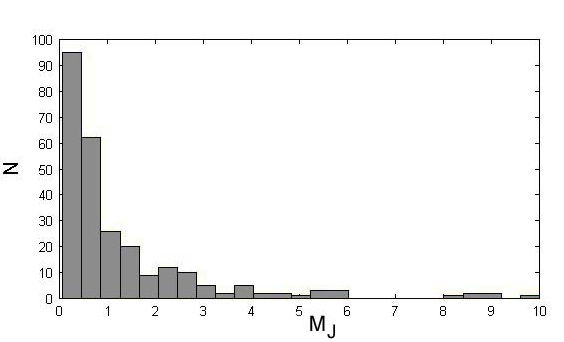}}
\caption{{\it Left}: Distribution of average radii as measured for all globulettes found in the Carina Nebula expressed in arcsec. {\it Middle}: The corresponding distribution of average radii expressed in kAU and adopting a distance to the complex of 2.9 kpc. The vertical arrow marks the peak in the corresponding accumulated size distribution for objects in seven \ion{H}{ii} regions (De Marco et al. \cite{mar06}). {\it Right}: Distribution of masses for Carina globulettes less massive than 10~M$_{J}$, expressed in Jupiter masses.}
\label{distribution}
\end{figure*}

\section{Results}
\label{sec:results}

We have found a total of 288 globulettes in the HST-images of the Carina complex.  Most of the objects are dark without any bright rims or halos, just like those found in surveys of other \ion{H}{ii} regions. The globulettes are spread over the entire region, but are more abundant along the western part of the V-shaped dark cloud and in areas surrounding Tr 14 and 16.  Examples of quite isolated globulettes can be found in Fields 10 and 25 in Figs.~\ref{fields2} and ~\ref{fields4}. Clusters of globulettes are found in, for example, Fields 12 and 41 in Figs.~\ref{fields2} and \ref{fields6}. The total number of globulettes found exceeds the number found in any other \ion{H}{ii} region. This large complex is comparatively well covered by HST observations and the number per unit area is comparable to the areas studied by De Marco et al. (\cite{mar06}).

\subsection{Distributions of radii and masses}  
\label{sec:distribute}

\begin{figure*}[t]
\centering
\includegraphics[angle=00, width=16cm]{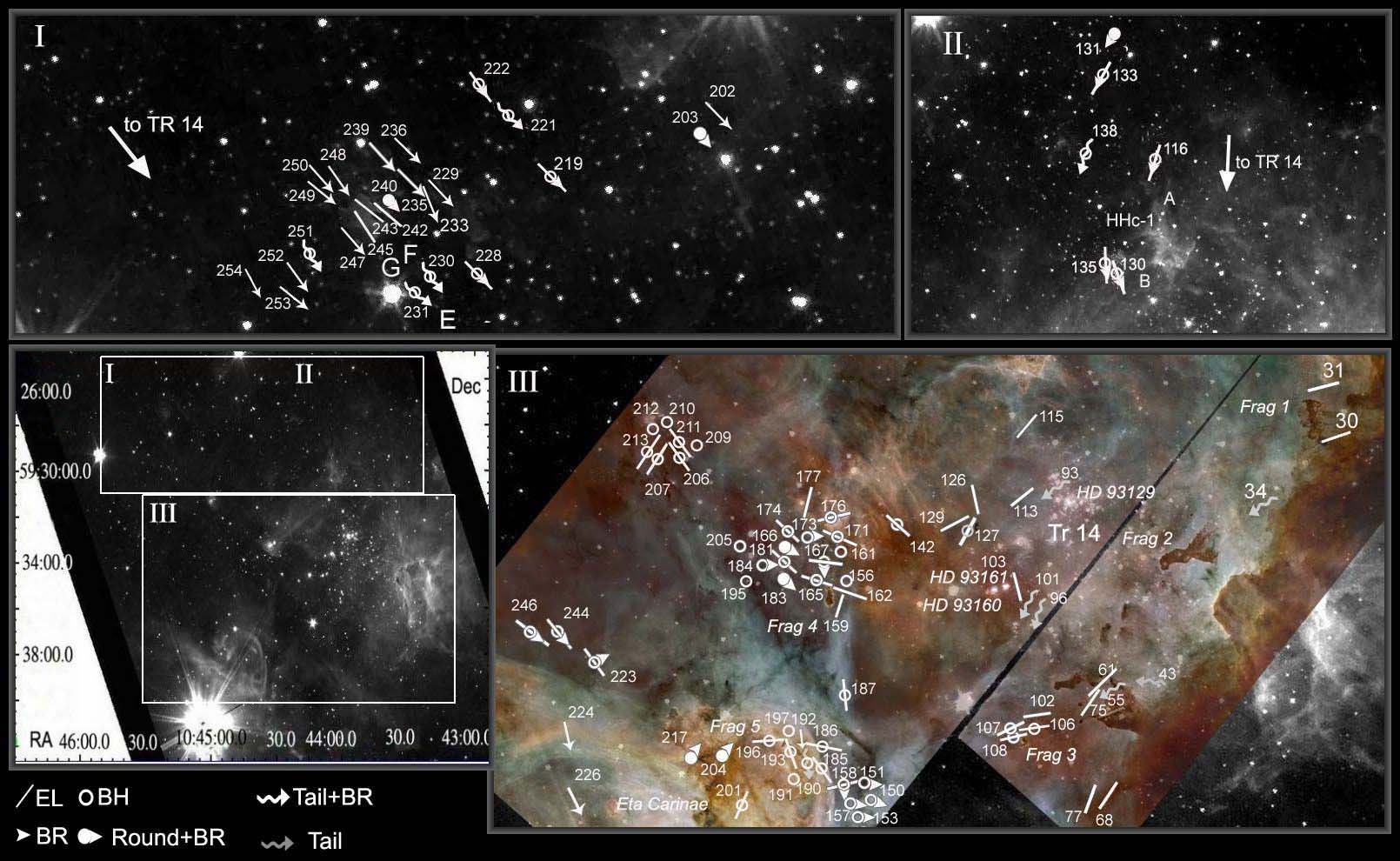}
\caption{In these charts the directions of the major axes of elongated objects are depicted in three areas, labelled I to III. In addition, the directions of objects with tails or bright rims present in round or elongated objects are indicated. The locations of these areas are shown in the lower-left panel on a map composed from Spitzer observations at 4.5 $\mu$m, This  background is used in all areas, and in area III  (lower-right) the optical image is superimposed as well. The symbols used are explained in bottom-left corner. North is up and east to the left in these images, and CN numbers are marked according to Tables A.1-A.5.}
\label{PAfields}
\end{figure*}

The left panel in Fig.~\ref{distribution} shows the distribution of average radii of the Carina globulettes expressed in arcsec, and in kAU in the middle panel. The bulk of the Carina globulettes have radii $<$~1000 AU, and the distribution increases steeply towards the detection limit. Hence, the Carina globulettes are, on the whole, significantly smaller than the accumulated distribution for the seven \ion{H}{ii} regions investigated by De Marco et al. (\cite{mar06}), which peaks at 2.5 kAU, and with detection limits similar to ours. We note that if we instead assume a distance of 2.3 kpc to the Carina complex, as advocated by Smith (\cite{smi08}), then the Carina globulettes would be even smaller by $\sim$~20\%.

The masses derived for the tiny globulettes in the Carina complex are consequently also, on the whole, considerably smaller than for other regions. Most of the globulettes have masses well within the domain of planetary masses. The right panel in Fig.~\ref{distribution} shows that the number of such objects increases rapidly below 3~M$_{J}$ towards the detection limit.  Only 4~\% of the Carina globulettes are more massive than 10~M$_{J}$, the most massive being CN~78 and 80 with $\sim$~130~M$_{J}$. This is in sharp contrast to the corresponding distribution in the Rosette Nebula that hosts a large number of more massive clumps, some with masses of several hundred M$_{J}$ (Paper~1). However, even though this complex is at half the distance to the Carina complex,  tiny objects with masses $<$~2~M$_{J}$ escape detection in this ground-based survey. The largest objects with masses  $>$~20~M$_{J}$ are located close to and along the V-shaped dust feature (Fields 9, 11, and 12) and to the south in Field 3, Position 30 (see Fig.~\ref{map}). They may represent relatively recent detachments from the nearby shell structures. 

The Carina globulettes not only differ in size from those in other regions, but also in density. Their average density amounts to $\rho$ = 2.8 10$^{-19}$ g cm$^{-3}$ compared to $\rho$~=~6.2~10$^{-20}$ g cm$^{-3}$ for those in the Rosette Nebula. In terms of number densities of molecular hydrogen they exceed 10$^{5}$ cm$^{-3}$ in several Carina globulettes.

\subsection{Orientations}
\label{sec:PAs}

Elongated globulettes, with or without tails, line up in the same direction in certain areas, but are more randomly oriented in others. The lower left-panel in Fig.~\ref{PAfields} shows the location of  three selected areas, I to III, projected on a strip composed from images obtained from Spitzer/IRAC 4.5~$\mu$m images (key no 23695360). The two upper panels show areas I and II on the 4.5~$\mu$m background, while for area III (bottom-right panel) the optical and Spitzer images are superimposed to better illustrate the locations of stars and bright and dark nebulosity. Included are also objects with pronounced bright rims and tails and a few round objects surrounded by bright halos that are distinctly brighter at one side. Obviously, some objects classified as round may in fact be elongated if they are oriented closer to the line of sight, and objects surrounded by bright halos flag the presence of bright rims on the remote side.

All of the objects depicted in areas I and II are oriented in about the same direction and point at the cluster Tr 14, located in area III, and it is clear that the objects have been sculpted by the interaction with photons coming from the bright stars in this cluster. It should be noted that there are also a large number of round, dark objects in these areas. In area III, elongated objects are more randomly oriented, particularly in the central part of the image. An example is CN~93 with 60~M$_{J}$ in Field 18, an isolated globulette seen in projection against the cluster Tr~14. However, both the direction of the tail and the bright rim indicates that the globulette is influenced by some object east of the cluster core. Globulettes just above and along the western extension of the V-shaped dust lane (in the right part of this panel), as well as a group to the left in the panel point in the general direction of the bright nebulosity surrounding $\eta$ Carina and Tr 16. Several O stars are spread over this nebulosity, and it is likely that their combined radiation caused the shaping and orientation of these elongated globulettes. In addition, we found the same predominant direction of objects in regions south of area III (not shown here).

\subsection{Peculiar objects} 
\label{sec:pec}

The Carina Nebula hosts a large number of clouds of very irregular shape like dark worm-like filaments and larger fragments, such as the "Defiant Fingers" described by Smith et al. (\cite{smi04}). Some of these fragments are accompanied by smaller cloudlets with irregular shapes that most likely have eroded from the larger ones. We did not include these objects in the list of globulettes. In Figs.~B1-B6 we have marked some larger fragments, since they might be important birth-places of globulettes (see Sect.~\ref{sec:disc}). 

\begin{figure}[t] 
\centering
\resizebox{8cm}{!}{\includegraphics[angle=00]{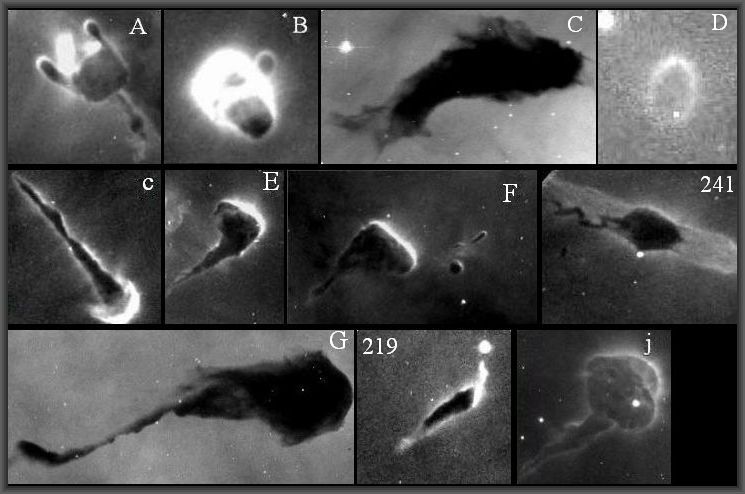}}
\caption{Examples of objects with peculiar shapes. The objects are marked in Figs. B.1-B.6, except for object 
$c$ which falls outside areas containing globulettes. This object and  $j$ are denoted in Smith et al. (\cite{smi03}). The images span: A 5\arcsec x 5\arcsec, B 4\arcsec x 4\arcsec, C 20\arcsec x 30\arcsec, D 3\arcsec x 3\arcsec, $c$ 8\arcsec x 8\arcsec, E 7\arcsec x 10\arcsec, F 20\arcsec x 10\arcsec, CN241 7\arcsec x 5\arcsec, G 35\arcsec x 18\arcsec, CN219 7\arcsec x 7\arcsec, $j$  15\arcsec x 15\arcsec. }
\label{strange}
\end{figure}

The border-line between what we define as a globulette or not is a bit arbitrary, but this is less important in our statistical analysis. Figure~\ref{strange} shows some examples of cloudlets with peculiar shapes, and their locations can be found in Figs.~B1-B6, except for $j$ which is in an HST field void of globulettes. This object, like object $c$, was noted and labelled in Smith et al. (\cite{smi03}). Objects B and D could be similar to standard globulettes, but since B appears to be surrounded by strong H$\alpha$ emission, and D is in the background behind strong foreground emission, these objects could not be measured for extinction. The other objects have peculiar dusty tails, and object A even shows three such outgrowths. These tails can be the result of erosion of elongated objects with peculiar density structure, or they may be examples of dust-enshrouded jets from embedded sources. Objects CN~219 and 241 were recognized as HH-objetcs in Smith et al. (\cite{smi03}), but their nature remains unclear. The objects CN 219 has a bright rim with a detached bright spot just to the north and the thin, twisted dust-tail in CN~241 is remarkable as is the surrounding, faint, and very elongated bright halo. Some of the peculiar objects in Fig.~\ref{strange} deserve a closer inspection.  

\begin{figure}[t]
\centering
\includegraphics[angle=00, width=10cm]{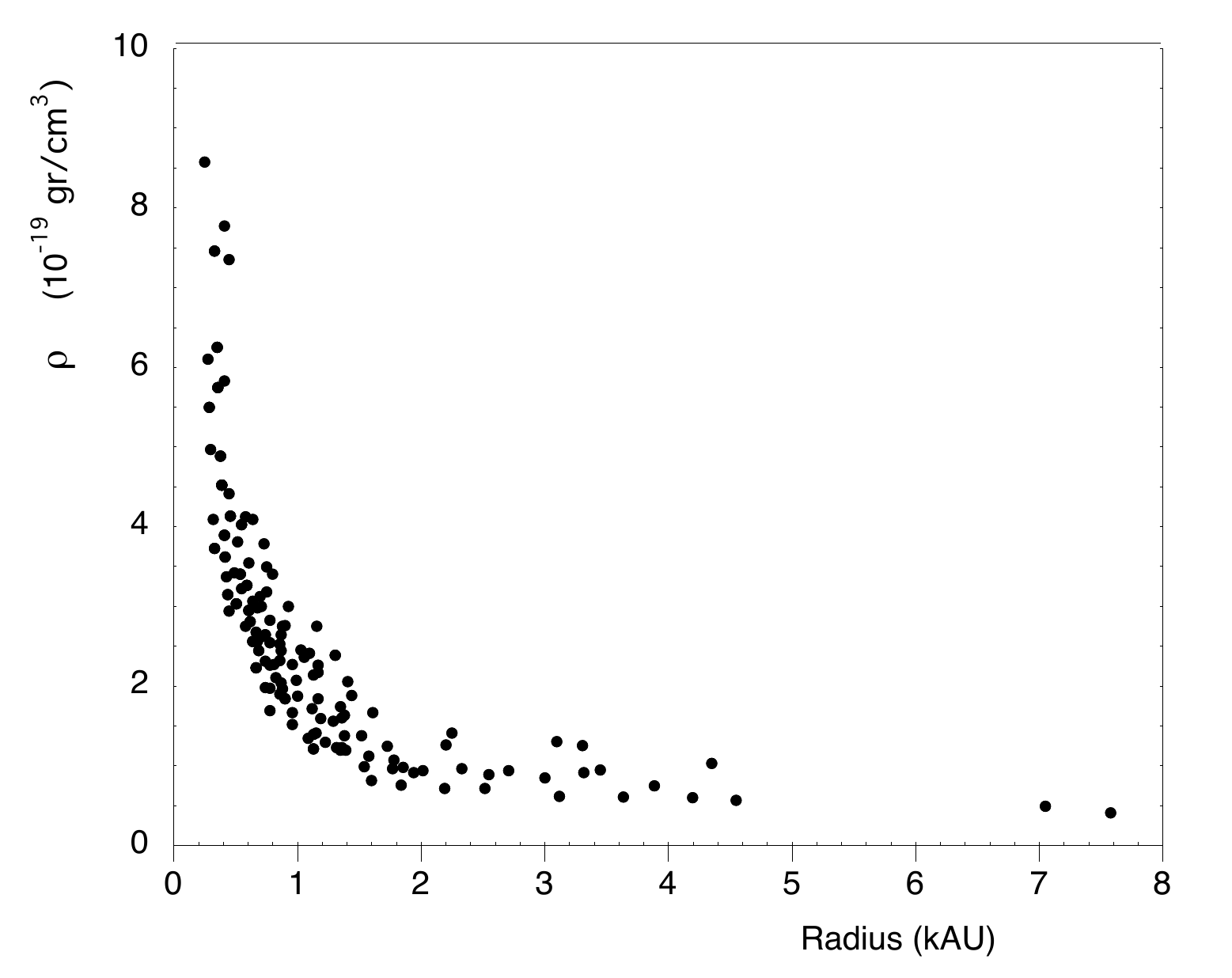}
\caption{Average density versus radius for all dark globulettes in the Carina complex.}
\label{RhoR}
\end{figure}

\section{Discussion}
\label{sec:disc}

None of the 288 globulettes coincides in position with any of the YSO candidates listed in Povich et al. (\cite{pov11}) and Gaczkowski et al. (\cite{gac13}). There are stars seen inside the boundaries of two globulettes in optical images, namely CN 138 (Fig.~\ref{mosaik}) and the object designated $j$ in Fig. 5 in Smith et al. (\cite{smi03}). These stars are likely to be foreground stars, since they show no sign of IR excess judging from the Two Micron All Sky Survey  (Skrutskie et al. \cite{skr06}) or existing Spitzer images. There are no obvious proplyd candidates in our sample, except possibly for CN 219 with a jet-like extension (see Figure~\ref{strange}) and listed as a Herbig-Haro object (HH 1011) in Smith et al. (\cite{smi10a}).  

\begin{figure*}[t] 
\centering
\resizebox{5.72cm}{!}{\includegraphics[angle=00]{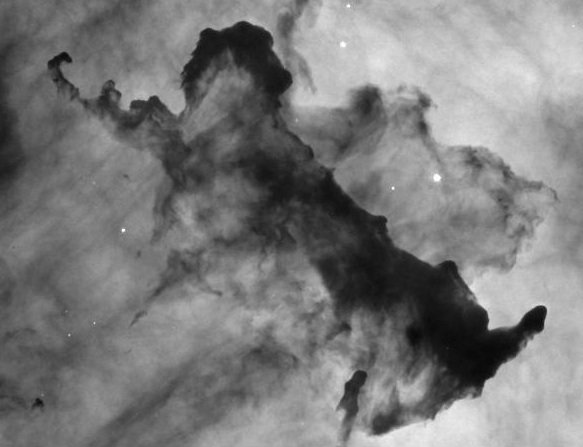}}
\resizebox{5.31cm}{!}{\includegraphics[angle=00]{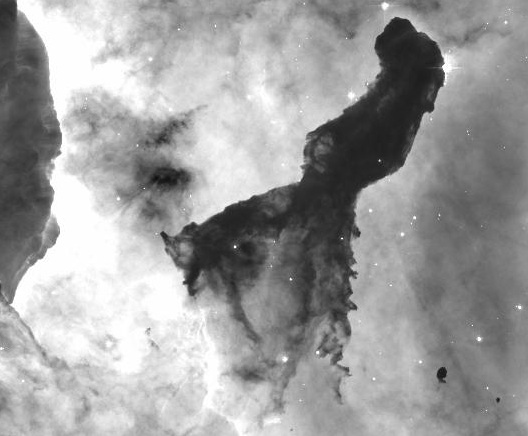}}
\resizebox{4.55cm}{!}{\includegraphics[angle=00]{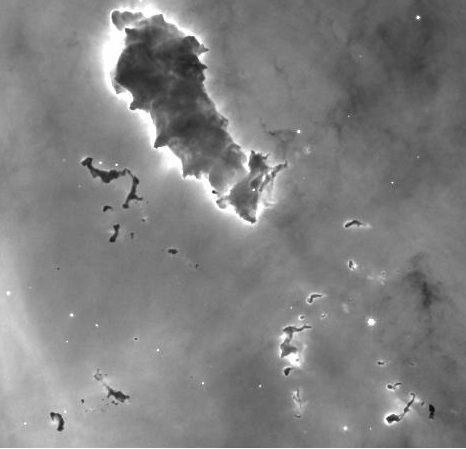}}
\caption{Three detached fragments in the Carina Nebula. From left to right: Fragments 1, 2, and 4 in Fields 7, 9, and 24 in Figs. B.1-B.6. These larger blocks contain denser cores, some of which appear to be detaching (see Frag. 1). Fragment 4 is surrounded by smaller cloudlets of different shapes. }
\label{fragments}
\end{figure*} 

The fraction of objects with bright rims and halos is 39\%, which is large compared to findings from other \ion{H}{ii} regions (De Marco et al. \cite{mar06}; Paper~1). In the central parts of the Carina Nebula objects with bright rims even dominate over those without indicating that the interaction with the radiation field is more intense closer to the centre.

An important finding is the statistical relation between average density and radius shown in Fig.~\ref{RhoR}, where we have selected only distinctly dark objects without bright rims and halos.  The smallest objects with radii $<$~1~kAU are on average four times denser than those with radii $>$2~kAU. The corresponding distribution including all objects in Tables 
A1-A5 shows the same general trend but with a larger scatter, and where the densities of BRs and BHs are systematically lower than the distribution in Fig.~\ref{RhoR}. This is expected since the masses for these objects are lower limits as pointed out in Sect.~\ref{sec:data}.  It is likely that the distribution in Fig.~\ref{RhoR} reflects how globulettes evolve with time.

\subsection{Origin and fate} 
\label{sec:origin}

Most dark formations seen in the optical images are located in front of the central regions of bright nebulosity. For the mass estimates we have considered two extreme cases: all residual emission in the darkest parts is due to foreground emission, or there is (practically) no foreground emission. The Carina region is very complex, and it is difficult to judge how deeply embedded a given globulette is in the warm nebulosity. For the same reason it is hard to determine the geometrical distance between a given globulette and other dust formations or clusters in the area. Some globulettes are quite isolated with projected distances to the closest dust complexes of more than 1.5~pc, for example the group of globulettes in Field 14 containing Tr 14. A possible scenario is that this group is the remnant of a larger cloud that gradually eroded in the intense radiation field from Tr 14. 

It was inferred in Paper~1 that globulettes in the Rosette Nebula originate from condensations in elephant trunks and shell features. In the Carina complex there are a number of isolated, larger fragments that must have detached from shell structures long ago. Such fragments are marked in Fields 7, 9, 12, 24, and 29. Figure~\ref{fragments} shows examples of such fragments and all contain condensations with masses similar to those found in globulettes. Fragment 4 is surrounded by a cluster of smaller irregular fragments that appear to be leftovers from a presumably larger block that once eroded. We note that Fragment 2, which hosts several condensations, looks like a detached elephant trunk composed by a network of thin twisted filaments similar to the threaded elephant trunks discussed in Carlqvist et al. (\cite{car03}).     

As discussed in Paper~2, the lack of distinct bright rims in H$\alpha$ may be traced to a combination of several circumstances. One is that the density distributions are rather flat and that the density is high even close to the surface where the gas is in molecular form as flagged by fluorescent H$_{2}$ emission. In addition, thin P$\beta$ emitting rims were discovered to be present in several objects, rims that in some cases appear to extend over the remote side of the objects. Such thin bright rims, not detected in H$\alpha$, were also found in several much smaller globulettes in the Rosette Nebula (M\"akel\"a et al. \cite{mak14}). Moreover, the Carina objects could be located at considerable distances from the bright UV-radiating stars, in which case the flux of exciting Lyman continuum photons is moderate producing only weak photodissociation in the outer layer at the dense surface.

The large number of quite isolated globulettes indicates that the objects have survived for a long time in the nebula. Most of these objects are tiny and dense, and unlike the larger objects they lack thin envelopes. It appears that the population of tiny globulettes in the Carina complex are in a more evolved state than those encountered in other \ion{H}{ii} regions, either because they have eroded faster in the intense radiation field, or because they are, on the whole, older. A likely scenario is that globulettes detach from larger molecular blocks, like shell structures, pillars, and fragments. Thin envelopes would gradually be lost with time, and the remnant cores may become denser with time. This scenario is further supported by the findings presented in Fig.~\ref{RhoR} showing that the average density is inversely proportional to radius. 

In Paper~1 we applied simple virial arguments to conclude that most globulettes in the Rosette Nebula could be gravitationally unstable, especially after considering the influence of an outer pressure from the surrounding warm plasma. When applying the same analysis to the Carina globulettes we found that the globulettes are close to virial equilibrium but none is bound, even when assuming an outer pressure (thermal plus turbulence) of the same magnitude as in the Rosette complex. On the other hand, the radiation pressure exerted by light from the numerous O stars in Carina should be much higher. This pressure acts on one side of the globulettes. The derived masses are subject to uncertainties (see Sect.~\ref{sec:obs}), and we note that very little extra mass is needed to confine the objects, as would be the case if they contain denser yet unresolved cores, or more speculatively, even Jupiter-sized planets. This would clarify why the globulettes appear to have survived for such a long time, as can be seen by their distribution over the nebula, where many objects are quite isolated and reside far away from larger molecular structures.

The total number of unbound planets in the Milky Way could amount to several hundred billion (Sumi et~al. \cite{sumi11}). Globulettes in  \ion{H}{ii} regions may be an additional source of such free-floating planetary-mass objects besides an origin in  circumstellar protoplanetary disks from where they are ejected (e.g. Veras et~al. \cite{veras09}).

\section{Conclusions}
\label{sec:conclude}

We have made an inventory of globulettes in the Carina Nebula complex based on existing HST narrow-band H$\alpha$ images. A total of 288 globulettes were listed and measured for size, mass, and density. Most objects are either round or slightly elongated, and many of the latter are oriented in the direction of massive young clusters in the area. We discuss why only a minority have developed bright H$\alpha$ emitting rims and/or tails, and we note that there is no evidence so far of any embedded young stars.

The Carina globulettes are, on the whole, much smaller and less massive than those recognized from HST surveys of a number of other \ion{H}{ii} regions. Practically all are of planetary mass, and most have masses less than one Jupiter mass. The corresponding mean densities are much higher than in other regions, exceeding number densities of 10$^{5}$ cm$^{-3}$ in several objects. We found a statistical relation between average density and size in the sense that the smallest globulettes are also the densest. Globulettes may detach from larger blocks of molecular gas, like isolated fragments, elephant trunks, and shell structures, after which their thinner envelopes evaporate and leave denser cores, which may become even more compressed with time.

From virial arguments we conclude that the objects are not bound unless they contain a bit more mass than inferred from the derived mean mass. Most of the tiny objects are quite isolated and located at projected distances of $>$~1.5 pc from the closest larger molecular structures, which indicates that the objects can survive for long times in the nebula. We speculate that the objects might contain denser cores or even planetary-mass objects that already have formed in their interior. 

We suggest that the Carina globulettes are a more evolved state than the larger and less dense objects that are abundant in other \ion{H}{ii} regions. Globulettes in  \ion{H}{ii} regions may be one source of the large number of free-floating planetary-mass objects that has been estimated to exist in the Galaxy.

\begin{acknowledgements}

We thank the referee Bo Reipurth for valuable comments and suggestions. This work was supported by the Magnus Bergvall Foundation, the L\"angmanska Kulturfonden, and the Swedish National Space Board.

\end{acknowledgements}

\newpage

\begin{appendix}

\section{Properties of the globulettes in the Carina complex}
\label{AppendixA}

\begin{table*}
\centering
\caption{List of globulettes in the Carina Nebula complex (the symbols are described in Section~\ref{sec:data}).}
\begin{tabular} {lccccccccccl}
 \hline\hline
     \noalign{\smallskip}
CN & Field & x & y & R.A. & Dec. & $\alpha$ & $\beta$ & P.A. & $\bar r $ & Mass & Remarks\\
&&&& (J2000.0) &(J2000.0) & (arcsec) &  (arcsec) & (degr.)  & (kAU) & (M$_{J}$) & \\
\noalign{\smallskip}
\hline
 \\

1   &  F1  & 360   & 4020   & 10:41:13.3 & -59:49:00  & 0.46 & 0.50    &    &   1.39    & 2.4    &    \\                                
2   &  F1  & 660	 & 1084   & 10:41:18.9 & -59:46:38  & 0.30 & 0.36    &    &   0.96    & 1.1    &     \\
3   &  F1  & 1792  & 3251   & 10:41:23.6 & -59:48:35   & 0.38 & 0.40   &    &   1.13    & 1.3    &    \\
4   &  F1  & 1654  & 378    & 10:41:26.2 & -59:46:13   & 0.18 & 0.19   &    &   0.54    & 0.4    &    \\
5   &  F1  & 1819  & 395    & 10:41:27.2 & -59:46:15   & 0.19 & 0.27   &    &   0.67    & 0.6    &    \\
6   &  F1  & 2029  & 221    & 10:41:28.8 & -59:46:09   & 0.26 & 0.33   & 23   &   0.86    & 1.2    &    T  \\
7   &  F2  & 670	 & 1907   & 10:41:28.9 & -59:45:53   & 0.23 & 0.25   & -43 &   0.69    &\it 0.6 &    BR,T \\
8   &  F1  & 2391  & 1935   & 10:41:29.1 & -59:47:36   & 0.20 & 0.22   &    &   0.61    & 0.5    &     \\
9   &  F1  & 2156  & 597    & 10:41:29.2 & -59:46:29   & 0.29 & 0.33   &    &   0.90    & 1.0    &    \\
10  &  F1  & 2752  & 1904   & 10:41:31.5 & -59:47:38   & 0.54 & 1.46   & 8 &   2.90    & \it 8.8    &  BR,EL,T\\
11  &  F1  & 2604  & 373	 & 10:41:32.4 & -59:46:22   & 0.31 & 0.51   & -38 &   1.19    & 2.0    &   EL  \\
12  &  F2  & 1207  & 1745   & 10:41:32.6 & -59:45:50   & 0.31 & 0.35   & -15   &   0.96    & 1.5    &  T \\
13  &  F1  & 2808  & 667    & 10:41:33.4 & -59:46:38   & 0.36 & 0.70   & 13 &   1.54    & 2.7    &  EL  \\
14  &  F2  & 1694  & 1606   & 10:41:35.9 & -59:45:48   & 0.39 & 0.46   &    &   1.23    & 1.8    &     \\
15  &  F3  & 2659  & 851	 & 10:41:44.9 & -59:43:44   & 1.47 & 1.67   &    &   4.55    & 40     &     \\
16  &  F3  & 2806  & 1494   & 10:41:45.2 & -59:44:17   & 0.52 & 0.71   &    &   1.78    & 4.5    &     \\
17  &  F3  & 2706  & 964    & 10:41:45.2 & -59:43:50   & 0.20 & 0.21   &    &   0.59    & 0.5    &     \\
18  &  F3  & 2972  & 933	 & 10:41:46.9 & -59:43:51   & 0.69 & 1.82   & 12&   3.64    & 22    &  EL,T  \\
19  &  F3  & 3527  & 1546   & 10:41:49.8 & -59:44:26   & 0.31 & 0.44   &    &   1.09    & 1.3     &     \\
20  &  F3  & 3656  & 1501   & 10:41:50.7 & -59:44:25   & 0.36 & 0.67   & 6 &   1.49    & \it 2.5    & BR,EL \\
21  &  F2  & 3781  & 272    & 10:41:51.2 & -59:45:02   & 0.64 & 1.00   & 10&   2.38    & \it 5.6    &  BR,EL \\ 
22  &  F4  & 2772  & 3493   & 10:42:20.7 & -59:19:45   & 0.35 & 0.40   &    &   1.09    & \it 1.6    &   BR   \\                    
23  &  F5  & 2719  & 1339   & 10:42:43.7 & -59:20:16   & 0.18 & 0.20   &    &   0.55    & 0.4    &      \\
24  &  F5  & 1880  & 938    & 10:42:48.7 & -59:20:42   & 0.35 & 0.72   & -5 &   1.55    & \it 2.8    &  BR,EL   \\               
25  &  F5  & 1242  & 524    & 10:42:52.4 & -59:21:08   & 0.25 & 0.47   & -3	&   1.04    & \it 1.5      & BR,EL    \\          
26  &  F8  & 1029  & 1467 & 10:43:15.9  & -59:28:07	& 0.13 & 0.14   &	   &   0.39    & 0.2   &      \\
27  &  F6  & 3938  & 1861  & 10:43:16.1 & -59:28:31	& 0.52 & 1.22   & 33 &   2.52    & 8.5   &  EL,T,C  \\
28  &  F8  & 973   & 1754   & 10:43:16.8    & -59:28:19	& 0.21 & 0.27	 &	   &   0.70    & 0.8   	&        \\
29  &  F6  & 2697  &	3970   & 10:43:17.6 & -59:30:33 & 0.26 & 0.28	 &	   &   0.78    & 0.9   &        \\
30  &  F7  & 1960  &	1480   & 10:43:19.9	 & -59:31:45 & 0.24 & 0.38	 & 20 &   0.90    & 1.5	   &     EL   \\
31  &  F7  & 3105  &	479    & 10:43:22.1	 & -59:30:31 & 0.36 & 0.61	 & 23 &   1.41    & 4.3	&     EL   \\
32  &  F7  & 2177  &	3163   & 10:43:27.5	 & -59:32:47 & 0.35 & 0.41	 &	   & 	 1.10    & 2.4   &        \\
33  &  F13 & 3571  & 682    & 10:43:31.4	 & -59:55:20 & 0.49 & 0.75	 & -13 &   1.80    & \it 4.0	& BH,EL    \\
34  &  F7  & 2583  &	3776   & 10:43:32.0	 & -59:33:00 & 0.23 & 0.26	 &	36   &   0.71    & 0.7	&     T    \\
35  &  F7  & 2883  &	4143   & 10:43:35.1	 & -59:33:06 & 0.19 & 0.25	 &	   &   0.64    & 0.5	&         \\
36  &  F7  & 3181  &	3863   & 10:43:35.6	 & -59:32:46 & 0.23 & 0.24	 &	   &   0.68    & 0.7	&         \\
37  &  F7  & 3360  &	4002   & 10:43:37.1	 & -59:32:46 & 0.11 & 0.13	 &	   &   0.35    & 0.2	&         \\
38  &  F9 & 3329  &	499	 & 10:43:39.1	 & -59:32:57 & 0.95 & 1.33	 &	   &   3.31    & 34	&  S$e$     \\
39  &  F9 & 3552  &	393	 & 10:43:39.8	 & -59:32:46 & 0.39 & 0.51	 &	   &   1.31    & 4.0	&  S$e$     \\
40  &  F10 & 1537  &	3364   & 10:43:43.8	 & -59:31:24 & 0.19 & 0.21	 &	   &   0.58    & 0.6	&         \\
41  &  F11 & 1933  &	2248   & 10:43:46.4	 & -59:36:05 & 0.18 & 0.25	 &	   &   0.62    & 0.5	&        \\
42  &  F11 & 1935  &	2353   & 10:43:46.8	 & -59:36:09 & 0.35 & 0.46	 &	   &   1.17    &2.6	&       \\
43  &  F11 & 2066  &	2364   & 10:43:47.6	 & -59:36:06 & 0.26 & 0.34	 &	-10 &   0.87    & 1.3	&      T \\
44  &  F9 & 2972  &	3165   & 10:43:47.6	 & -59:34:56 & 0.33 & 0.40	 &	   &   1.06    & 2.1	&       \\
45  &  F9 & 3255  &	2797   & 10:43:47.7	 & -59:34:33 & 0.13 & 0.13	 &	   &   0.38    & 0.2	&       \\
46  &  F9 & 3325  &	2858   & 10:43:48.3	 & -59:34:33 & 0.11 & 0.13	 &	   &   0.35    & 0.2    &       \\
47  &  F9 & 2914  &	3633   &	10:43:49.1 & -59:35:16	& 0.13 & 0.14	 &	   &   0.39    & 0.2	&        \\
48  &  F9 & 2893  &	3668   &	10:43:49.1 & -59:35:18	& 0.11 & 0.12	 &	   &   0.33    & 0.1	&       \\
49  &  F10 & 2378  &	3919   &	10:43:50.4 & -59:31:18	& 0.36 & 0.44	 &	   &   1.16    & 3.2	   &       \\
50  &  F9 & 3229  &	3794   &	10:43:51.4 & -59:35:14	& 0.30 & 0.41	 &	   &   1.03    & 2.0	   &       \\
51  &  F12 & 506  & 3746   &	10:43:52.5 & -59:39:03	& 0.56 & 0.78	 &	   &   1.94    & 5.0	&       \\
52  &  F16 & 3116  &	1570   &	10:43:52.9 & -59:58:49	& 0.27 & 0.34	 &	   &   0.88    & 1.0	&       \\
53  &  F11 & 2489  &	3191   &	10:43:53.0 & -59:36:27	& 0.21 & 0.30	 &	   &   0.74    & 0.8	&       \\
54  &  F11 & 2530  &	3187   &	10:43:53.2 & -59:36:25	& 0.33 & 0.48	 &	   &   1.17    & 2.7	   & C to 56  \\
55  &  F11 & 2573  &	3152   &	10:43:53.3 & -59:36:23	& 0.12 & 0.16	 & 39	 & 	 0.41    & 0.3	&      T  \\
56  &  F11 & 2554  &	3196   &	10:43:53.3 & -59:36:25	& 0.26 & 0.35	 &	   & 	 0.88    & 1.4	&  C to 54  \\
57  &  F11 & 2589  &	3155   &	10:43:53.4   & -59:36:22 & 0.12 & 0.13	 &	   & 	 0.36    & 0.2	&   \\
58  &  F11 & 2873  &	2777	 & 10:43:53.5	 & -59:35:59	& 0.25 & 0.29	 &	   & 	 0.78    & 1.0	&       \\
59  &  F11 & 2611  &	3162   &	10:43:53.6 & -59:36:22	& 0.13 & 0.15	 &	   & 	 0.41    & 0.2	&       \\

\hline

\end{tabular}
\label{globCSD} 
\end{table*} 

\begin{table*}
\centering
\caption{List of globulettes in the Carina Nebula complex. }
\begin{tabular} {lccccccccccl}
 \hline\hline
     \noalign{\smallskip}
CN & Field & x & y & R.A. & Dec. & $\alpha$ & $\beta$ & P.A. & $\bar r $ & Mass & Remarks\\
&&&& (J2000.0) &(J2000.0) & (arcsec) &  (arcsec) & (degr.)  & (kAU) & (M$_{J}$) & \\
\noalign{\smallskip}
\hline
 \\  
60  &  F11 & 2952  &	2770	 & 10:43:53.9	 & -59:35:56	& 0.15 & 0.17	 &	  	&   0.46    & 0.3	&       \\
61  &  F11 & 2569  &	3315   &	10:43:53.9 & -59:36:29	& 0.27 & 0.70	 & -6 &   1.38    & 2.7	&     EL \\ 
62  &  F11 & 2874  &	2959   &	10:43:54.2 & -59:36:06	& 0.16 & 0.24	 &	  	&   0.58    & 0.4	&        \\
63  &  F11 & 2722  &	3170   &	10:43:54.2 & -59:36:19	& 0.13 & 0.14	 &	  	&   0.39    & 0.2	&       \\
64  &  F12 & 830   &	3731   &	10:43:54.2 & -59:38:53	& 0.44 & 0.47	 &	  	&   1.32    & 2.1	   &       \\
65  &  F11 & 2595  &	3360   &	10:43:54.2 & -59:36:31	& 0.15 & 0.16	 &	  	&   0.45    & 0.2	&      \\
66  &  F16 & 2341  &	102	 & 10:43:54.5	 & -60:00:11	& 0.33 & 0.41	 &	  	&   1.07    & 2.2	&      \\
67  &  F12 & 891	 & 3782   & 10:43:54.7  & -59:38:54 & 0.17 & 0.18	 &	  	&   0.51    & 0.3	&      \\
68  &  F12 & 827	 & 3883   & 10:43:54.8 & -59:38:59 	& 0.75 & 1.12	 & -10 &   2.71    & 14	&  EL    \\
69  &  F12 & 1124  &	3501   &	10:43:54.9 & -59:38:35	& 0.36 & 0.42	 &	  	&   1.13    & 1.5	&       \\
70  &  F16 & 2304  &	190	 & 10:43:54.9	 & -60:00:08	& 0.42 & 0.57	 &	  	&   1.44    & 4.2	&      \\
71  &  F11 & 2766  &	3381   &	10:43:55.2 & -59:36:26	& 0.13 & 0.16	 &	  	&   0.42    & 0.2	&       \\
72  &  F12 & 1251  &	3451   &	10:43:55.4 & -59:38:30	& 0.39 & 0.50	 &	  	&   1.29    & 2.5	&       \\
73  &  F11 & 2833  &	3366   &	10:43:55.5 & -59:36:24	& 0.12 & 0.16	 &	  	&   0.41    & 0.2	&       \\
74  &  F11 & 2790  &	3454   &	10:43:55.6 & -59:36:29	& 0.20 & 0.23	 &	  	&   0.62    & 0.5	&       \\
75  &  F11 & 2796  &	3536   &	10:43:55.9 & -59:36:32	& 0.18 & 0.33	 & -9 &   0.74    & 0.7	&     EL \\
76  &  F11 & 2805  &	3527   &	10:43:56.0 & -59:36:31	& 0.18 & 0.26	 &	  	&   0.64    & 0.5	&        \\
77  &  F12 & 1049 &	3910   &	10:43:56.1   & -59:38:54 & 0.73 & 1.34   & -16 &   3.00    & 17     &   EL \\
78  &  F11 & 2606  &	3914   &	10:43:56.4 & -59:36:53	& 2.06 & 2.80	 &	   &	 7.05    & 128	&  C  \\
79  &  F17 & 1970  &	196	 & 10:43:56.4	 & -59:39:03	& 1.16 & 1.22	 &	  	&   3.45    & 29	&       \\
80  &  F11 & 2844  &	3613   &	10:43:56.6 & -59:36:33	& 2.30 & 2.93	 &	   &	 7.58    & 132	&       \\
81  &  F17 & 2057  &	149    &	10:43:56.7 & -59:38:58	& 0.63 & 0.65	 &	  	&   1.86    & 4.7	&       \\
82  &  F14 & 2761  &	894	 & 10:43:56.7	 & -59:31:32	& 0.22 & 0.28	 &	  	&   0.73    & 1.1	&       \\
83  &  F12 & 1222  &	3852   &	10:43:56.8 & -59:38:47	& 0.34 & 0.44	 &	   &   1.13    & 1.3	&       \\
84  &  F14 & 2839  &	813	 & 10:43:56.8	 & -59:31:26	& 0.13 & 0.15	 &	  	&   0.41    & 0.2	&     \\
85  &  F12 & 1406  & 3654   &	10:43:56.9 & -59:38:33	& 0.70 & 0.82	 &	  	&   2.20    & 10	&       \\
86  &  F17 & 2182  &	172	 & 10:43:57.5	 & -59:38:56	& 0.44 & 0.50	 &	  	&   1.36    & 3.0	&       \\
87  &  F12 & 1314  &	3927   &	10:43:57.5 & -59:38:47	& 0.71 & 0.84	 &	  	&   2.25    & 12	&       \\
88  &  F14 & 3211  &	691	 & 10:43:58.1	 & -59:31:10	& 0.13 & 0.21	 &	  	&   0.49    & 0.3	&       \\
89  &  F11 & 3786  &	2786	 & 10:43:58.4	 & -59:35:32	& 0.13 & 0.15	 &	  	&   0.41    & 0.2	&       \\
90  &  F17 & 2303  &	274    &	10:43:58.5 & -59:38:56	& 0.58 & 0.81	 &	  	&   2.02    & 5.8	&    \\
91  &  F14 & 3214  &	863    &	10:43:58.9 & -59:31:16	& 0.12 & 0.12	 & 	&   0.35    & 0.2	&        \\
92  &  F14 & 2726  &	1454   &	10:43:58.9 & -59:31:54	& 0.22 & 0.27	 & 	&   0.71    & 0.8	&       \\
93  &  F18 & 2146  &	258	 & 10:43:59.6	 & -59:32:38     & 1.31 & 1.69	 &	5  &   4.35    & 63	& T,S$i$ \\
94 &  F18 & 384   &	3025   & 10:44:01.1	 & -59:35:22	& 0.14 & 0.16	 & 	&   0.44    & 0.2	&       \\
95 &  F14 & 2903  &	1866   & 10:44:01.6	 & -59:32:04	& 0.09 & 0.11	 & 	&   0.29    & 0.1	&       \\
96 &  F18 & 575  &	2990   &	10:44:01.9 & -59:35:15	& 0.18 & 0.26	 &   -40	&   0.64    & 0.8	   &    T \\
97 &  F14 & 2890  &	2055   &	10:44:02.4 & -59:32:12	& 0.12 & 0.13	 &	  	&   0.36    & 0.2	&      \\
98 &  F14 & 2865  &	2085   &	10:44:02.4 & -59:32:14	& 0.37 & 0.46	 &	  	&   1.20    & \it 3.2	&  BR,S$f$  \\
99 &  F18 & 559   &	3151   &	10:44:02.5 & -59:35:22	& 0.20 & 0.22	 &	  	&   0.61    & 0.5	&       \\
100 &  F14 & 2994  & 2059   &	10:44:02.8 & -59:32:08	& 0.36 & 0.54	 &	  	&   1.31    & \it 4.0	& S$f$    \\
101 &  F18 & 660   &	3153   &	10:44:03.0 & -59:35:19	& 0.17 & 0.30	 &  -13  &   0.68    & 0.6	&    EL,T  \\
102 &  F12 & 3489  & 2374   &	10:44:03.1 & -59:36:40	& 0.12 & 0.23	 & 3 &   0.51    & 0.3	   &     EL  \\
103 &  F18 & 722   &	3146   &	10:44:03.3 & -59:35:17	& 0.15 & 0.23   &  11 &   0.55    & 0.5	&   EL,T  \\
104 &  F18 & 708   & 3183   & 10:44:03.4	 & -59:35:19	& 0.17 & 0.21	 &	   &   0.55    & 0.5	&   \\
105 &  F12 & 3420  & 2634   & 10:44:03.8  & -59:36:53	& 0.10 & 0.10	 &	   &   0.29    & 0.1	&  \\
106 &  F12 & 3392  & 2729   & 10:44:03.9  & -59:36:57	& 0.10 & 0.23	 & 27&   0.48    & \it 0.2	&  BH,EL\\
107 &  F12 & 3409  & 2742   & 10:44:04.1  & -59:36:57	& 0.10 & 0.16	 & 7 &   0.38    & \it 0.2	&    BH,EL \\
108 &  F12 & 3447  & 2703   & 10:44:04.2  & -59:36:55	& 0.10 & 0.17	 & 21&   0.38    & \it 0.2	 &  BH,EL\\
109 &  F12 & 3450  & 2789   & 10:44:04.5  & -59:36:58	& 0.12 & 0.19	 & 22&   0.45    & \it 0.2	 &   BH,EL  \\
110 &  F17 & 2657  & 1400   &	10:44:04.7 & -59:39:32	& 0.33 & 0.45	 &	   &   1.13    & 2.3	 &     \\
111 &  F17 & 2603  &1490   & 10:44:04.8	 & -59:39:37	& 0.49 & 0.62	 &	   &   1.61    & 5.2	 &      \\
112 &  F17 & 2587  & 1545   &	10:44:04.9 & -59:39:40	& 0.43 & 0.50	 &	   &   1.35    & 3.2	 &      \\
113 &  F14 & 2686  & 2944   &	10:44:05.2 & -59:32:52	& 0.26 & 0.69	 & 43  &   1.38    & 3.2	 &    EL  \\
114 &  F17 & 2737  & 1427   & 10:44:05.3	 & -59:39:30	& 0.11 & 0.10	 & 	&   0.30    & 0.1	&       \\
115 &  F14 & 3623  & 2009   &	10:44:05.8 & -59:31:46	& 0.14 & 0.24	 & 44 &   0.55    &  0.5	 &    EL  \\
116 &  F15 & 2760  & 2864   &	10:44:06.4 & -59:28:13	& 0.17 & 0.35	 & -31  & 0.75    &  \it 0.8	 &   BH,EL  \\
117 &  F14 & 2847  & 3192   &	10:44:07.0 & -59:32:56	& 0.26 & 0.38	 &	 	 &  0.93    &  1.8	 &      \\
118 &  F14 & 3190  & 2832   &	10:44:07.2 & -59:32:31	& 0.08 & 0.09	 &	 	 &  0.25    &  0.1	 &      \\
119 &  F17 & 2961  & 1710   &	10:44:07.6 & -59:39:35	& 1.22 & 1.68	 &	 	 &  4.20    &  33	 &       \\
\hline

\end{tabular}
\label{globCSD} 
\end{table*} 

\begin{table*}
\centering
\caption{List of globulettes in the Carina Nebula complex.}
\begin{tabular} {lccccccccccl}
 \hline\hline
     \noalign{\smallskip}
CN & Field & x & y & R.A. & Dec. & $\alpha$ & $\beta$ & P.A. & $\bar r $ & Mass & Remarks\\
&&&& (J2000.0) &(J2000.0) & (arcsec) &  (arcsec) & (degr.)  & (kAU) & (M$_{J}$) & \\
\noalign{\smallskip}
\hline
 \\  

120 & F14  & 3189  &	2936   &	10:44:07.6	 & -59:32:35	& 0.11 & 0.12	 &	 	 &  0.33    &  0.2	 &      \\
121 & F14  & 3203  &	3000   &	10:44:07.9	 & -59:32:37	& 0.13 & 0.14	 &	 	 &  0.39    &  0.2	 &      \\
122 & F14  & 2861  &	3513   &	10:44:08.5	 & -59:33:08	& 0.19 & 0.23	 &	 	 &  0.61    &  0.6	 &      \\
123 & F14  & 3194  & 	3190   &	10:44:08.7	 & -59:32:45	& 0.23 & 0.29	 &	 	 &  0.75    &  1.1	 &      \\
124 & F14  & 3272  &	3217   &	10:44:09.3	 & -59:32:43	& 0.18 & 0.23	 &	 	 &  0.59    &  0.5	 &      \\
125 & F14  & 2906  &	3805   &	10:44:09.9	 & -59:33:17	& 0.12 & 0.14	 &	 	 &  0.38    &  0.2	 &      \\
126 & F14  & 3011  &	3781   &	10:44:10.4   	& -59:33:13 	& 0.20 & 0.39	 & 16	 &  0.86    &  0.9	 &    EL  \\  
127 & F14  & 2991  &	3827   &	10:44:10.5   	& -59:33:16	& 0.28 & 0.49	 & -5  &  1.12    &  \it 2.3	 &   BH,EL \\  
128 & F14  & 2909  &	3956   &	10:44:10.6	 & -59:33:23	& 0.11 & 0.12	 &	 	 &  0.33    &  0.2	 &       \\
129 & F14  & 2977  &	3907   &	10:44:10.7	 & -59:33:19	& 0.15 & 0.39	 & 43	 &  0.78    &  0.8	 &   EL   \\
130 & F20  & 1107  &	2190   &	10:44:12.6	 & -59:31:03	& 0.19 & 0.29	 &	20 &  0.70    & \it 0.7	 &   BH,EL  \\
131 & F21  & 1458  &	1517   &	10:44:12.8	 & -59:25:53	& 0.17 & 0.21	 &	 	 &  0.55    &  \it 0.4	 &   BR  \\
132 & F21  & 1441  &	1599   &	10:44:13.0	 & -59:25:57	& 0.16 & 0.20	 &	 	 &  0.52    & \it  0.4	 &    \\
133 & F21  & 1347  &	1758   &	10:44:13.2	 & -59:26:06	& 0.38 & 0.89	 & -11  &  1.84    & \it 3.8	 &   BH,EL  \\
134 & F19  & 1781  &	2650   &	10:44:15.0	 & -59:35:41	& 0.21 & 0.23	 &		 &  0.64    &  0.6	 &       \\
135 & F20  & 1747  &	2070   & 10:44:15.3	 	& -59:30:38	& 0.13 & 0.23	 & 3   &  0.52    & \it 0.3	 &   BH,EL   \\
136 & F22  & 1814  &	1588   &	10:44:16.8	 & -59:42:36	& 0.10 & 0.14	 &		 &  0.35    & \it 0.2	 &   BH  \\
137 & F22  & 2295  &	1451   &	10:44:18.8	 & -59:42:16	& 0.41 & 0.52	 &	-26 &  1.35    & \it 2.5  	 &   BH,T   \\
138 & F20  & 3991  &	352	 & 10:44:19.1	 	& -59:28:20	& 0.32 & 0.67	 & -24 &  1.44    & \it 2.7	 & BH,EL,T  \\
139 & F22  & 2338  &	1531   &	10:44:19.4	 & -59:42:18	& 0.11 & 0.15	 &		 &  0.38    & \it 0.2	 &  BH   \\
140 & F28  & 287   	&	834	 & 10:44:19.9	 	& -59:33:12	& 0.09 & 0.10	 &		 &  0.28    &  0.1	 &     \\
141 & F28  & 245   &		1067   &	10:44:20.6	 & -59:33:23	& 0.13 & 0.16	 &		 &  0.42    &  0.2	 &     \\
142 & F28  & 439   &		1398   &	10:44:22.9	 & -59:33:30	& 0.41 & 0.90	 & -17 &  1.90    & \it 10	 &   BH,EL   \\
143 & F26  & 3864  &	1397   &	10:44:28.1	 & -60:00:08	& 0.18 & 0.26	 &		 &  0.64    &  0.5	 &     \\
144 & F26  & 3874  &	1486   &	10:44:28.1	 & -60:00:03	& 0.26 & 0.34	 &		 &  0.87    &  1.2	 &      \\
145 & F26  & 3875  &	1543   &	10:44:28.2	 & -60:00:00	& 0.51 & 0.71	 &		 &  1.77    &  4.0	 &      \\
146 & F26  & 3811  &	1412   &	10:44:28.5   	& -60:00:07	& 0.27 & 0.32	 &	    &  0.86    &  1.1 	 &      \\
147 & F26  & 3810  &	1434   &	10:44:28.5	 & -60:00:06	& 0.24 & 0.27	 &		 &  0.74    &  0.8	 &      \\
148 & F26  & 3759  & 	1382   & 10:44:28.8	 	& -60:00:09 	& 0.23 & 0.28   &     &  0.74    &  0.8  &    \\
149 & F26  & 3623  &	506	 & 10:44:28.9	 	& -60:00:53	& 0.25 & 0.32	 & 	 &  0.83    &  0.9	 &     \\
150 & F23  & 2896  &	2906   & 10:44:29.1	 	& -59:38:40	& 0.11 & 0.12	 &		 &  0.33    & \it 0.2	 &  BH    \\
151 & F23  & 3070  &	2960   &	10:44:30.2	 & -59:38:37	& 0.13 & 0.16	 &		 &  0.42    & \it 0.3	 &   BH  \\
152 & F28  & 552   &		3222   &	10:44:30.6	 & -59:34:41	& 0.17 & 0.25	 &		 &  0.61    &  0.5	 &      \\
153 & F23  & 2756  &	3560   &	10:44:31.1	 & -59:39:10	& 0.18 & 0.33	 &		 &  0.74    & \it 0.6	 &  BH    \\
154 & F28  & 503   &		3487   &	10:44:31.3	 & -59:34:53	& 0.12 & 0.13	 &		 &  0.36    &  0.2	 &      \\
155 & F26  & 3610  &	 3069   & 10:44:31.4   	& -59:58:47 	& 0.23 & 0.23   &     &  0.67    &  0.5  &C   \\
156 & F28  & 657   &		3347   &	10:44:31.6	 & -59:34:43	& 0.12 & 0.16	 &		 &  0.41    & \it 0.2	 &  BH    \\
157 & F23  & 2990  &	3452   &	10:44:31.9	 & -59:39:01	& 0.11 & 0.11	 &		 &  0.32    & \it 0.1	 &  BH    \\
158 & F23  & 3159  &	3340   &	10:44:32.2   	& -59:38:49	& 0.12 & 0.22	 & 24  &  0.49    & \it 0.3	 & BH,EL      \\
159 & F28  & 650   & 	3554   & 10:44:32.4   	& -59:34:51 	& 0.18 & 0.30   & -11 &  0.69    &  0.6  &  EL \\
160 & F27  & 3512  &	750	 & 10:44:32.5	 	& -59:57:30	& 0.42 & 1.19	 &  6	 &  2.33    &  9.1	 &EL,T  \\
161 & F28  & 1561  &	2363   &	10:44:32.6	 & -59:33:36	& 0.10 & 0.12	 &	  	 &  0.32    & \it 0.1	 &    BH   \\
162 & F28  & 671   & 	3584   & 10:44:32.6	 	& -59:34:52 	& 0.11 & 0.21   & 44  &  0.46    &  0.3  &  EL  \\
163 & F29  & 156   &		490	 & 10:44:32.6	 	& -59:37:36	& 0.15 & 0.20	 &	    &  0.51    & \it 0.5	 &    BH   \\
164 & F26  & 3436  & 	3148   & 10:44:32.6   	& -59:58:44	 & 0.24 & 0.23   &     &  0.68    &  0.7  & C  \\
165 & F28  & 731   & 	3521   & 10:44:32.7	 	& -59:34:47	 & 0.25 & 0.50   &  -5 &  1.09    & \it 1.6  & BH,EL \\
166 & F28  & 1180  &	2970   &	10:44:32.9	 & -59:34:12	& 0.11 & 0.15	 &	    &  0.38    &  \it 0.2	 &    BR   \\
167 & F28  & 960   &		3282   &	10:44:32.9 	 & -59:34:31	 & 0.26 & 0.47   &  -37	 &  1.06    & \it 1.3	 & BH,EL \\
168 & F27  & 3434  &	856	 & 10:44:33.1	 	& -59:57:25	& 0.14 & 0.16	 &	  	 &  0.44    &  0.2	 &      \\
169 & F26  & 2931  &	196	 & 10:44:33.1	 	& -60:01:14	& 0.34 & 0.43	 &	  	 &  1.12    &  1.8	 &     \\
170 & F26  & 3400  &	863	 & 10:44:33.3	 	& -59:57:25	& 0.11 & 0.12	 &	  	 &  0.33    &  0.1	 &      \\
171 & F27  & 989   &		3389	 & 10:44:33.5	 	& -59:34:34	 & 0.11 & 0.22	 & -13 &  0.48    & \it 0.2	 &  BH,EL   \\ 
172 & F26  & 2859  &	463	 & 10:44:33.9	 	& -60:01:01	& 0.50 & 0.55	 &	  	 &  1.52    &  3.6	 &       \\
173 & F28  & 1165  &	3251	 & 10:44:33.9	 	& -59:34:24	& 0.18 & 0.27	 &	    &  0.65    & \it 0.6	 &  BH    \\
174 & F28  & 1147  &	3298	 & 10:44:34.0	 	& -59:34:26	& 0.14 & 0.32	 &-4	 &  0.67    & \it 0.4	 & BH,EL\\
175 & F26  & 2899  &	971	 & 10:44:34.1	 	& -60:00:36	& 0.46 & 0.73	 &  2	 &  1.73    &  4.8	 & EL,T   \\
176 & F28  & 1231  & 	3217	 & 10:44:34.1	 	& -59:34:20	 & 0.12 & 0.24   &  21	 &  0.52    &\it  0.3	 &  BH,EL \\
177 & F28  & 1250  & 	3203	 & 10:44:34.2   		& -59:34:19	 & 0.31 & 0.78   &  -29 &  1.58    & 3.3	 &  EL \\
178 & F26  & 2733  &	168	 & 10:44:34.4	 	& -60:01:17	& 0.15 & 0.16	 & 	 &  0.45    &  0.3	 &       \\
179 & F29  & 455   &		586    & 10:44:34.5	 	& -59:37:30	& 0.13 & 0.15	 &	  	 &  0.41    &  0.4	 &       \\
\hline

\end{tabular}
\label{globCSD} 
\end{table*} 

\begin{table*}
\centering
\caption{List of globulettes in the Carina Nebula complex.}
\begin{tabular} {lccccccccccl}
 \hline\hline
     \noalign{\smallskip}
CN & Field & x & y & R.A. & Dec. & $\alpha$ & $\beta$ & P.A. & $\bar r $ & Mass & Remarks\\
&&&& (J2000.0) &(J2000.0) & (arcsec) &  (arcsec) & (degr.)  & (kAU) & (M$_{J}$) & \\
\noalign{\smallskip}
\hline
 \\  

180 & F28  & 1585  &	2904   & 10:44:34.8	 & -59:33:57	& 0.11 & 0.13	 &	  	 &  0.35    &  0.2	 &     BH  \\
181 & F28  & 1330  & 	3301	 & 10:44:35.0	 & -59:34:21     & 0.10 & 0.24    &  -9&  0.49    & \it 0.2	 & BH, EL \\
182 & F26  & 2599  &	499	 & 10:44:35.6	 & -60:01:01 	& 0.18 & 0.19	 &	    &  0.54    &  0.4	 &       \\
183 & F28  & 1340  &	3499   & 10:44:35.8	 & -59:34:29	& 0.12 & 0.15	 &	    &  0.39    & \it 0.3	 &    BR   \\
184 & F28  & 1707  &	3005   & 10:44:35.9	 & -59:33:58	& 0.19 & 0.26	 &	    &  0.65    & \it 0.6	 &    BH   \\
185 & F29  & 306   &		1298   &	10:44:36.8 & -59:38:03	& 0.15 & 0.34	 & 9  &  0.71    & \it 0.8	 &   BH,EL   \\
186 & F29  & 385   &		1214   &	10:44:36.8 & -59:37:57	& 0.12 & 0.19	 & -42   &  0.45    & \it 0.3	 &  BH,EL   \\
187 & F29  & 325   &		1329   &	10:44:37.0 & -59:38:03	& 0.13 & 0.22	 & -13  &  0.51    & \it 0.5	 &     BH,EL   \\
188 & F29  & 967   &		666    &	10:44:37.5 & -59:37:17	& 0.10 & 0.12	 &	    &  0.32    &  0.1	 &        \\
189 & F26  & 2389  &	1025   &	10:44:37.5 & -60:00:37	& 0.60 & 0.67	 &	    &  1.84    &  3.5	 &       \\
190 & F29  & 389   &		1433   &	10:44:37.8 & -59:38:05	& 0.18 & 0.34	 & 21 &  0.75    & \it 1.0	 &     BH,EL,T  \\
191 & F29  & 416   &		1430   &	10:44:37.9 & -59:38:04	& 0.09 & 0.10	 &	  	 &  0.28    & \it 0.1	 &   BH  \\
192 & F29  & 601   &		1217   & 10:44:37.9   & -59:37:50	& 0.13 & 0.17	 & 9   &  0.44    & \it 0.3	 & BR.EL  \\
193 & F29  & 466   &		1381   &	10:44:37.9 & -59:38:01	& 0.14 & 0.23	 & 7   &  0.54    & \it 0.3	 & BH,EL   \\
194 & F29  & 1203  &	581	 & 10:44:38.3   & -59:37:06	& 0.12 & 0.14	 &     &  0.38    &  0.2	 &     \\
195 & F28  & 2194  &	3098   &	10:44:38.8 & -59:33:47	& 0.09 & 0.10	 &	 	 &  0.29    & \it 0.1 &  BH    \\
196 & F29  & 763   & 	1319   & 10:44:39.2	 & -59:37:49	& 0.26 & 0.45	 & 6	 &  1.03    & \it 2.0	 &  BH,EL   \\
197 & F29  & 877   &		1190   & 10:44:39.2   &	-59:37:40 & 0.10 & 0.12	 &	 	 &  0.32    & \it 0.1	 &   BH  \\
198 & F29  & 1014  &	1365   & 10:44:40.6   &	-59:37:43 & 0.15 & 0.20	 &	 	 &  0.51    &  0.3	 &     \\
199 & F25/28  & 1945  & 3616   & 10:44:41.3   &	-59:46:26 & 1.22 & 1.46	 &	 	 &  3.89    &  33 &     \\
200 & F25  & 3596  &	1558   & 10:44:42.1   &	-59:44:14 & 0.35 & 0.59	 & 2  &  1.36    &  2.3	 & EL    \\
201 & F33  & 380   &		1177   & 10:44:45.2   &	-59:39:07 & 0.15 & 0.25	 & -26   &  0.58    & \it  0.4	 &   BH,EL   \\
202 & F32  & 1780  &	3807   & 10:44:46.8   &	-59:26:25 & 0.18 & 0.30	 & -1  &  0.70    & \it  0.7	 &   BR,EL   \\
203 & F32  & 1884  &	3764   & 10:44:47.0   &	-59:26:30 & 0.25 & 0.33	 &	 	 &  0.84    & \it  1.1	 &  BR   \\
204 & F33  & 1657  &	317	 & 10:44:48.1   &	-59:37:53 & 0.27 & 0.34	 &	 	 &  0.88    & \it  1.1	 &   BR \\
205 & F28  & 3952  &	3839   & 10:44:51.0   &	-59:33:25 & 0.11 & 0.10	 &	 	 &  0.30    & \it 0.1	 &  BH   \\
206 & F31  & 1558  &	3034   & 10:44:55.0   &	-59:32:03 & 0.23 & 0.37	 & 31 &  0.87    & \it 1.0	 &   BH,EL,C to 207 \\
207 & F31  & 1578  &	3041   & 10:44:55.2   &	-59:32:03 & 0.18 & 0.41 	 &  -7 &  0.81    & \it 0.8	 &BH,EL,C to 206  \\
208 & F28  & 2749  &	4107   & 10:44:55.5   &	-59:47:38 & 0.22 & 0.29	 &	 	 &  0.74    &  0.6	 &      \\
209 & F31  & 1729  &	2933   & 10:44:55.5   &	-59:31:54 & 0.12 & 0.17	 &	 	 &  0.42    & \it 0.2	 &  BH   \\
210 & F31  & 1744  &	2954   & 10:44:55.7   &	-59:31:54 & 0.18 & 0.24	 &	 	 &  0.61    & \it 0.5	 &   BH  \\
211 & F31  & 1776  &	2936   & 10:44:55.8   &	-59:31:53 & 0.15 & 0.28	 &  -8&  0.62    & \it 0.5	 &   BH,EL   \\
212 & F31  & 1822  &	2927   & 10:44:56.0   &	-59:31:51 & 0.11 & 0.13	 &	    &  0.35    & \it 0.2	 &    BH   \\
213 & F31  & 1839  &	2938   & 10:44:56.1   &	-59:31:51 & 0.24 & 0.50	 &  -10 &  1.07    & \it 1.3	 &    BH,EL  \\
214 & F33  & 2034  & 	1833   & 10:44:56.4   &	-59:38:40 & 0.23 & 0.29	 &	    &  0.75    &  1.0	 &     \\
215 & F30  & 2894  &	2278   & 10:44:57.1   &	-59:47:36 & 0.37 & 0.42	 &	    &  1.15    &  1.6	 &       \\
216 & F34  & 634   &		493	 & 10:44:58.1   &	-59:45:30 & 0.24 & 0.31	 &	    &  0.80    & \it 0.8	 &  BH    \\
217 & F33  & 2984  &	1488   & 10:44:59.7   &	-59:37:56 & 0.21 & 0.26	 &	    &  0.68    & \it 0.6	 &    BR   \\
218 & F33  & 3039  &	2209   & 10:45:03.0   &	-59:38:22 & 0.17 & 0.20	 &	    &  0.54    &  0.4	 &      \\
219 & F32  & 2175  &	991    & 10:45:05.0   &	-59:26:57 & 0.34 & 1.23	 & 44 &  2.28    & \it 4.0	 & BH,EL,HH 1011 \\
220 & F34  & 1926  &	694	 & 10:45:05.8   &	-59:45:00 & 0.27 & 0.33	 &	    &  0.87    &  1.0	 &      \\
221 & F32  & 1901  &	849	 & 10:45:06.0   &	-59:26:44 & 0.17 & 0.39	 & 36&  0.81    & \it 0.6	 & BH,EL,T \\
222 & F32  & 1800  & 	790	 & 10:45:06.5   &	-59:26:39 & 0.22 & 0.63	 & 35 &  1.22    & \it 1.4	 &  BH,EL   \\
223& F35    & 1157  & 	892	 & 10:45:07.0   &	-59:36:22 & 0.19 & 0.32	 & -4  &  0.74    & \it 0.8	 &  BH,EL    \\
224 & F33  & 3463  &	2708   & 10:45:07.3   &	-59:38:28 & 0.09 & 0.12	 & 9  &  0.30    & \it 0.1	 &  BR,EL  \\
225 & F33  & 3876  &	2237   & 10:45:07.4   &	-59:37:56 & 0.13 & 0.17	 &	    &  0.43    &  0.2	 &    \\
226 & F33  & 2807  & 	3660   & 10:45:08.0   &	-59:39:25 & 0.11 & 0.21	 &23 &  0.46    & \it 0.2	 &  BR,EL  \\
227 & F34  & 1339  &	2233   & 10:45:08.7   &	-59:46:19 & 0.83 & 0.93	 &		 &  2.55    & 11	 &     \\
228 & F36  & 3883  &	3930   & 10:45:10.2   &	-59:28:18 & 0.12 & 0.31	 & 41&	 0.62    & \it 0.4	 &     BH,EL \\
229 & F36  & 2492  &	3336   & 10:45:14.8   &	-59:27:11 & 0.26 & 0.45	 & 41  &  1.03    & \it 1.5	 &  BR,EL  \\
230 & F36  & 3665  &	3179   & 10:45:15.2   &	-59:28:10 & 0.23 & 0.94	 & 31 &  1.70    & \it 1.0	 & BH,EL,T  \\
231 & F36  & 3991  &	3108   & 10:45:15.5   &	-59:28:26 & 0.17 & 0.30	 & -8  &  0.68    & \it 0.4	 &   BH,EL,T  \\
232 & F34  & 2953  &	1788   & 10:45:15.5   &	-59:45:14 & 0.45 & 0.65	 &     &  1.60    &  2.5	 &    \\
233    & F36  & 3003  &	3106   & 10:45:16.0   &	-59:27:37 & 0.10 & 0.19	 & 35  &  0.42    & \it 0.2	 &  BR,EL  \\
234  & F34  & 3018  &	1878   & 10:45:16.2   &	-59:45:15 & 0.53 & 0.98	 & -14  &  2.19    &  5.6	 &   EL  \\
235     & F36  & 2999  &	3076   & 10:45:16.2   &	-59:27:37 & 0.11 & 0.23	 & 35  &  0.49    & \it 0.3	 &  BR,EL \\
236  & F36  & 2080  &	3045   & 10:45:16.9   &	-59:26:52 & 0.12 & 0.21	 & -4  &  0.49    & \it 0.3	 &  BR,EL  \\
237    & F36  & 2850  &	2967   & 10:45:17.0   &	-59:27:03 & 0.09 & 0.18	 & 36  &  0.39    & \it 0.2	 &  BR,EL  \\
238    & F36  & 2053  &	2993   & 10:45:17.3   &	-59:26:50 & 0.11 & 0.18	 & 37  &  0.42    & \it 0.2	 &  BR,EL \\
239  & F36  & 2086  &	2949   & 10:45:17.6   &	-59:26:52 & 0.19 & 0.34	 & 38  &  0.77    & \it 1.0	 & BR,EL    \\
\hline

\end{tabular}
\label{globCSD} 
\end{table*} 

\begin{table*}
\centering
\caption{List of globulettes in the Carina Nebula complex.}
\begin{tabular} {lccccccccccl}
 \hline\hline
     \noalign{\smallskip}
CN & Field & x & y & R.A. & Dec. & $\alpha$ & $\beta$ & P.A. & $\bar r $ & Mass & Remarks\\
&&&& (J2000.0) &(J2000.0) & (arcsec) &  (arcsec) & (degr.)  & (kAU) & (M$_{J}$) & \\
\noalign{\smallskip}
\hline
 \\  

240  & F36  & 3557  &	2638   & 10:45:18.8   &	-59:28:07 & 0.43 & 0.49	 &  	 &  1.33    &\it  2.3	 &   BR  \\
241  & F34  & 4031  &	1315   & 10:45:19.4   &	-59:44:23 & 0.86 & 1.29	 & -15  &  3.12    &  14	    & EL,T,HH 900  \\
242  & F36  & 3013  &	2526   & 10:45:19.8   &	-59:27:40 & 0.11 & 0.20	 & -5  &  0.45    &  0.5	 &  EL  \\
243  & F36  & 2815  &	2471   & 10:45:20.3   &	-59:27:31 & 0.22 & 0.33	 & -8  &	 0.80    &  1.3	 &   EL  \\
244  & F35   & 3833  &	803    & 10:45:20.3   &	-59:34:54 & 0.19 & 0.56	 & -6 &	 1.09    & \it 1.9	 &  BH,EL  \\
245      & F36  & 2956  &	2448   & 10:45:20.4   &	-59:27:38 & 0.12 & 0.20	 & -4  &  0.46    & \it 0.3	 &  EL  \\
246      & F35   & 3925  &	862    & 10:45:21.0   &	-59:34:53 & 0.11 & 0.22	 &-21 &	 0.48    & \it 0.4	 &   BH,EL \\
247  & F36  & 2947  &	2037   & 10:45:23.1   &	-59:27:39 & 0.17 & 0.35	 & 41 &	 0.75    & \it 0.5	 &  BR,EL \\
248  & F36  & 2438  &	2035   & 10:45:23.4   &	-59:27:14 & 0.15 & 0.26 	 & 34 &  0.59    & \it 0.5	 &   BR,EL  \\
249  & F36  & 2642  &	2011   & 10:45:23.4   &	-59:27:24 & 0.14 & 0.25	 &-6  &  0.56    & \it 0.4	 &  BR,EL   \\
250 & F36  & 2426  &	1969   & 10:45:23.8   &	-59:27:13 & 0.10 & 0.20	 & 41 &  0.44    & \it 0.2	 &   BR,EL, \\
251  & F36  & 3344  &	1606   & 10:45:25.7   &	-59:28:01 & 0.11 & 0.19	 & 44 &  0.43    & \it 0.2	 &  BH,EL,T  \\
252  & F36  & 3446  &	1332   & 10:45:27.4   &	-59:28:07 & 0.12 & 0.41	 & 36 &  0.77    & \it 0.4	 & BR,EL \\
253  & F36  & 3855  &	1259   & 10:45:27.6   &	-59:28:28 & 0.20 & 0.57	 & -7 &  1.12    & \it 0.9	 &   BR,EL  \\
254  & F36  & 3465  &	722	 & 10:45:31.4   & 	-59:28:10 & 0.10 & 0.25	 & 28 &  0.51    & \it 0.2	 & BR,EL \\
255  & F37  &1063  &	1293  & 10:45:43.3   &	-59:41:48 & 0.17 & 0.32	 & -8  &  0.71    & \it 0.5	 &  BH,EL  \\
256  & F37  & 1218  &	1373   & 10:45:49.3   &	-59:41:20 & 0.19 & 0.22	 &     &  0.59    &  0.5	 &   \\
257  & F37  & 2123  &	1452   & 10:45:49.6   &	-59:41:23 & 0.24 & 0.32	 &     &  0.81    &  0.9	 & T  \\
258  & F38  & 1950  & 	772    & 10:45:52.4   &	 -60:09:22 & 0.94 & 1.20   &     &  3.10    &  29     &\\
259  & F39  & 2083  &	1920   & 10:45:56.4   &	-60:06:50 & 0.25 & 0.37	 &	    &  0.90    &  1.0	 &     \\
260  & F39  & 1727  &	1822   & 10:45:58.5   &	-60:07:00 & 0.39 & 0.54	 &	    &  1.35    &  2.2	 &     C  \\
261  & F40  & 3160  &	2494   & 10:46:25.4   &	-60:04:46 & 0.23 & 0.34	 & 24    &  0.83    & \it 0.8	 &  BH,T  \\
262  & F40  & 1860  &	859    & 10:46:32.5   &	-60:06:16 & 0.17 & 0.52	 & 11   &  1.00    & \it 0.8	 &  BH,EL  \\
263  & F40  & 1829  &	2265   & 10:46:34.0   &	-60:05:07 & 0.22 & 0.74	 &  13 &  1.39    & \it 1.4	 &   BH,EL  \\
264  & F40  & 1684  &	2025   & 10:46:34.8   &	-60:05:20 & 0.18 & 0.39	 & 13  &  0.83    & \it 0.9	 & BH,EL   \\
265  & F40  & 1595  &	3374   & 10:46:36.6   &	-60:04:13 & 0.29 & 0.35	 &	    &  0.93    & \it 0.9	 &  BH,C \\
266  & F40  & 1099  &	2480   & 10:46:39.1   &	-60:05:01 & 0.18 & 0.36	 & 6   &  0.78    & \it 0.6	 &   BH,EL\\
267  & F40  & 1183  &	3446   & 10:46:39.4   &	-60:04:13 & 0.18 & 0.28	 &	 	 &  0.67    & \it 0.4	 &  BH   \\
268  & F40  & 548   &	2567   & 10:46:42.8   &	-60:05:01 & 0.12 & 0.21	 &  10 &  0.48    & \it 0.3	 &    BH,EL \\
269  & F40  & 572   &	3295   & 10:46:43.3   &	-60:04:24 & 0.22 & 0.35	 &  27 &  0.83    & \it 0.7	 &  BH,EL \\
270  & F40  & 536   &	3228   & 10:46:43.5   &	-60:04:28 & 0.15 & 0.31	 &  24 &  0.67    & \it 0.5	 & BH,EL  \\
271  & F41  & 2754  &	3454   & 10:46:44.0   &	-60:08:16 & 0.35 & 0.64	 &  -38 &  1.44    & \it 2.0	 &   BH,EL \\
272  & F42  & 4047  &	547	 & 10:46:44.2   &	-60:07:29 & 0.23 & 0.41	 &	 -43	 &  0.93    & \it 0.6	 &  BH,EL  \\
273  & F41  & 2206  &	3203   & 10:46:47.4   &	-60:08:33 & 0.25 & 0.29	 &	 	 &  0.78    &  0.6	 &    \\
274  & F41  & 1750  &	609	 & 10:46:48.0   &	-60:10:44 & 0.18 & 0.28	 &  40 &  0.67    &  0.5	 &   EL  \\
275  & F41  & 2193  &	3922   & 10:46:48.2   &	-60:07:57 & 0.40 & 0.68	 &  -25 &  1.7    & \it 3.3	 &  BH,EL  \\
276  & F41  & 2118  & 	3648   & 10:46:48.4   &	-60:08:11 & 0.25 & 0.32	 &	 	 &  0.83    & \it 0.7	 &   BR  \\
277  & F41  & 1974  &	3499   & 10:46:49.2   & -60:08:20 & 0.32 & 0.34	 &	 	 &  0.96    &  1.0	 &     \\
278  & F41  & 1870  &	3499   & 10:46:49.9   & -60:08:20 & 0.60 & 0.91	 & 32  &  2.19    & \it 5.3	 &    BH,EL \\
279  & F41  & 1788  &	3611   & 10:46:50.6   & -60:08:15 & 0.37 & 0.39	 &	 	 &  1.10     & \it 1.4	 &  BH     \\
280  & F41  & 1664  & 	2864   & 10:46:50.7   & -60:08:53 & 0.25 & 0.29	 &	 	 &  0.78    &  0.7	 &     \\
281  & F43  & 3479  &	1455   & 10:46:51.4   & -60:03:35 & 0.34 & 0.47	 &   7&  1.17    & \it 2.9	 &  BR, EL,T  \\
282  & F41  & 1472  &	3476   & 10:46:52.5   & -60:08:24 & 0.20 & 0.26	 &	 	 &  0.67    &  0.5	 &      \\
283  & F43  & 2949  &	1487   & 10:46:54.9   & -60:03:37 & 0.31 & 0.50	 &  26&  1.17    &  2.2	 &  EL,C to 268  \\
284  & F43  & 2933  &	1486   & 10:46:55.0   & -60:03:38 & 0.26 & 0.42	 &  15&  0.99    &  1.5	 &  EL,C to 267  \\
285  & F43  & 2818  &	1519   & 10:46:55.8   & -60:03:37 & 0.13 & 0.14	 &     &  0.39    &  0.2	 &    \\
286  & F42  & 2371  &	3032   & 10:46:57.7   & -60:05:38 & 0.82 & 1.47	 &  35&  3.32    & 25	 &  EL    \\
287  & F43  & 2498  &	3796   & 10:47:00.0   & -60:01:46 & 0.25 & 0.45	 &   -23&  1.02    & \it 1.3	 & BH,EL   \\
288 & F43   & 2411  & 	4092   & 10:47:00.9   & -60:01:32 & 0.32 & 0.37	 &	 	 &  1.00    & 1.4	    &    \\
\hline 
                                                                                                                              
\end{tabular}
\label{globCSD} 
\end{table*} 

\section{Atlas of fields with globulettes.}
\label{AppendixB}

\begin{figure*}[t] 
\centering
\resizebox{6.9cm}{!}{\includegraphics[angle=00]{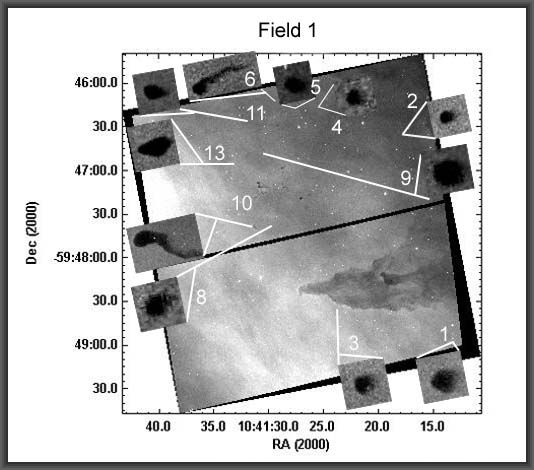}}
\resizebox{6.9cm}{!}{\includegraphics[angle=00]{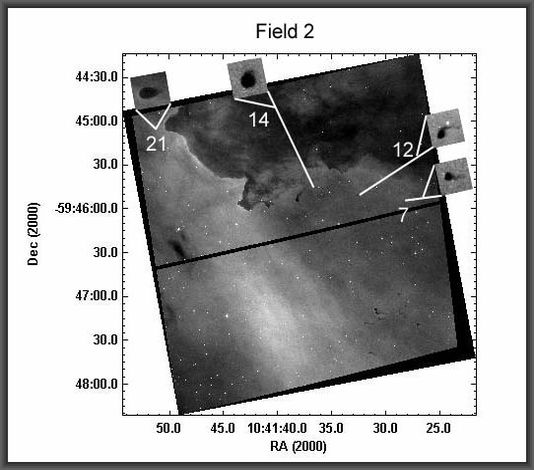}}
\resizebox{6.9cm}{!}{\includegraphics[angle=00]{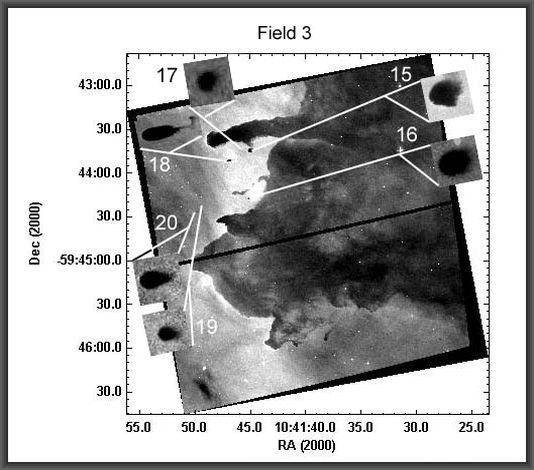}}
\resizebox{6.9cm}{!}{\includegraphics[angle=00]{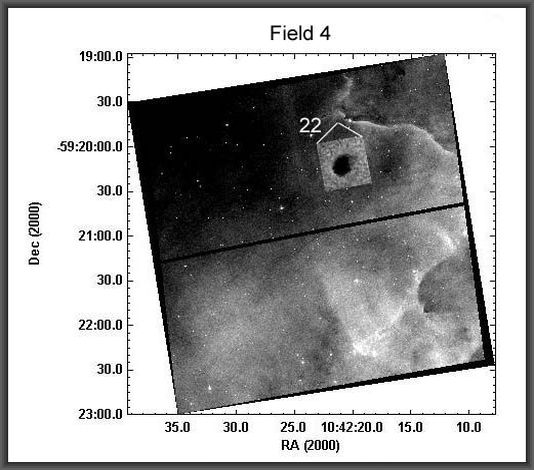}}
\resizebox{6.9cm}{!}{\includegraphics[angle=00]{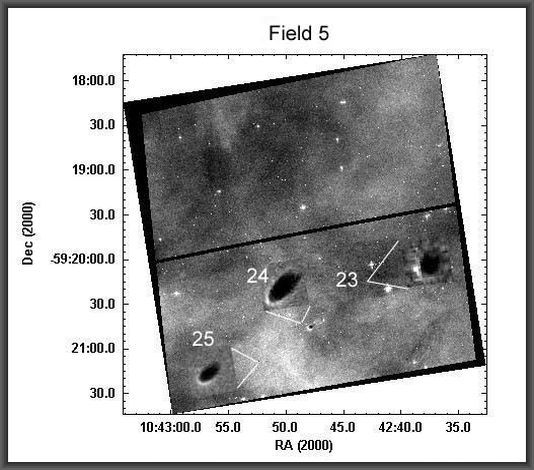}}
\resizebox{6.9cm}{!}{\includegraphics[angle=00]{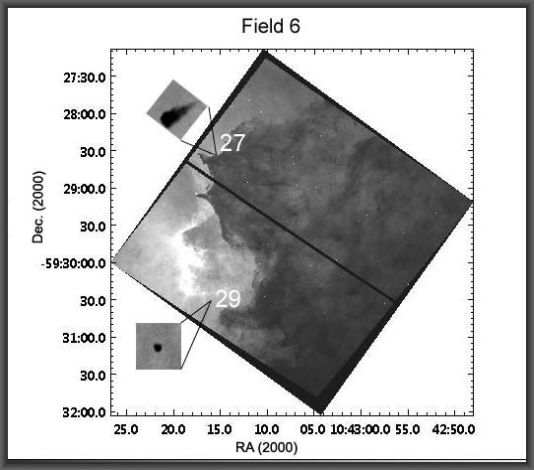}}
\resizebox{6.9cm}{!}{\includegraphics[angle=00]{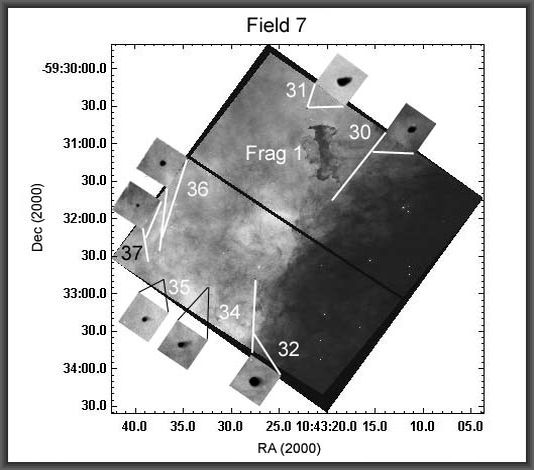}}
\resizebox{6.9cm}{!}{\includegraphics[angle=00]{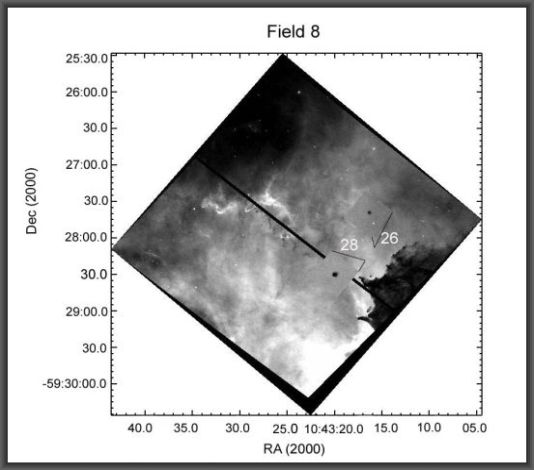}}
\caption{Fields 1 to 8 with object designation and inserted enlargements of globulettes.}
\label{fields1}
\end{figure*} 

\begin{figure*}[t] 
\centering
\resizebox{7cm}{!}{\includegraphics[angle=00]{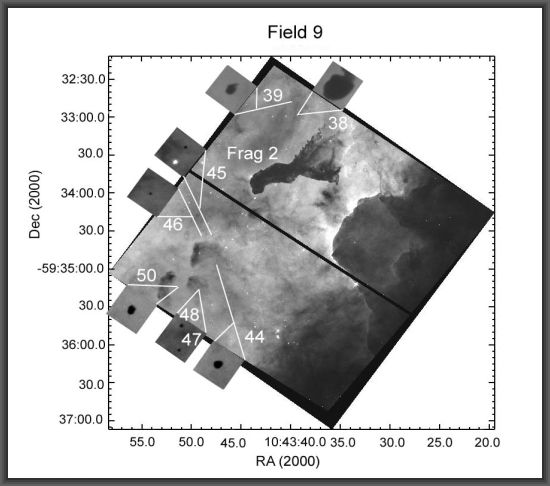}}
\resizebox{7cm}{!}{\includegraphics[angle=00]{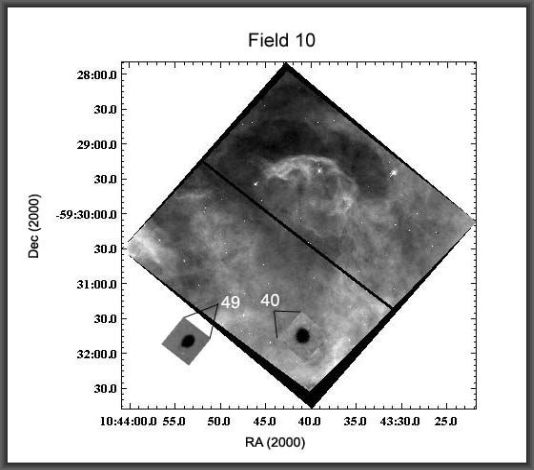}}
\resizebox{7cm}{!}{\includegraphics[angle=00]{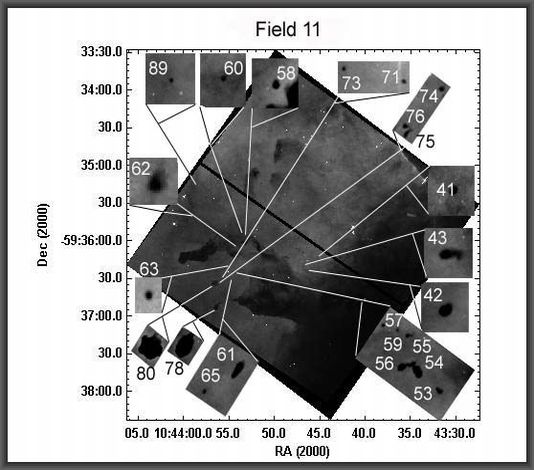}}
\resizebox{7cm}{!}{\includegraphics[angle=00]{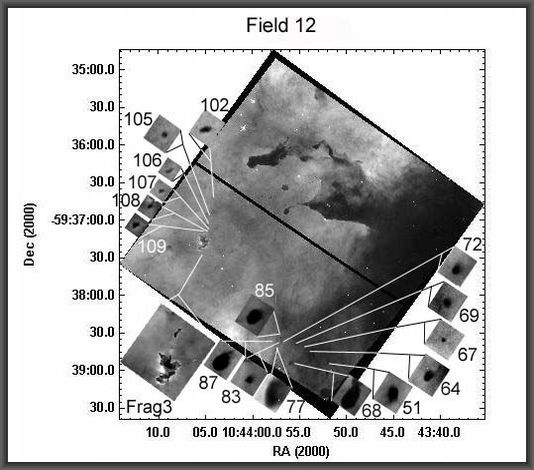}}
\resizebox{7cm}{!}{\includegraphics[angle=00]{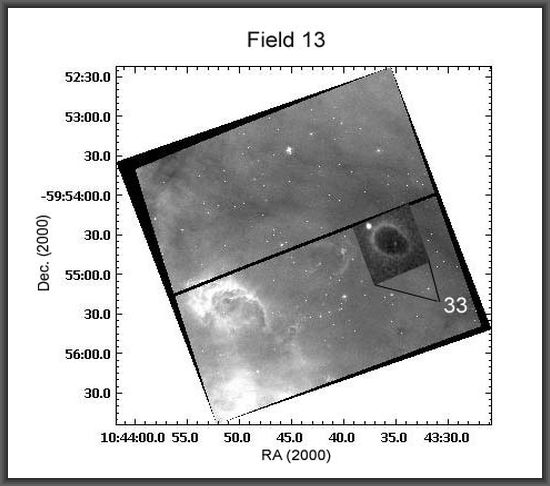}}
\resizebox{7cm}{!}{\includegraphics[angle=00]{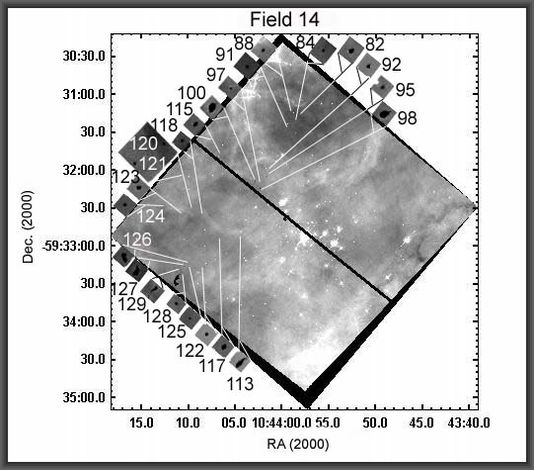}}
\resizebox{7cm}{!}{\includegraphics[angle=00]{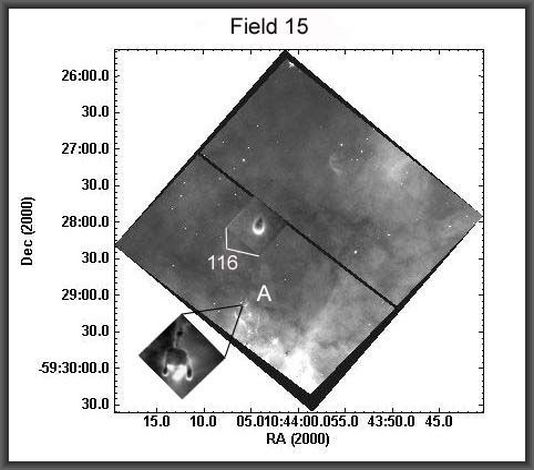}}
\resizebox{7cm}{!}{\includegraphics[angle=00]{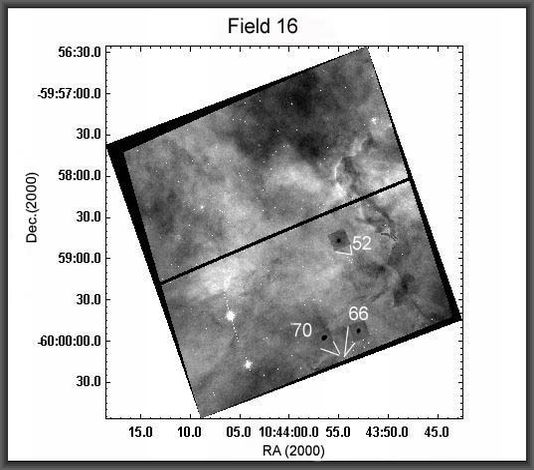}}
\caption{Fields 9 to 16.}
\label{fields2}
\end{figure*} 

\begin{figure*}[t] 
\centering
\resizebox{7cm}{!}{\includegraphics[angle=00]{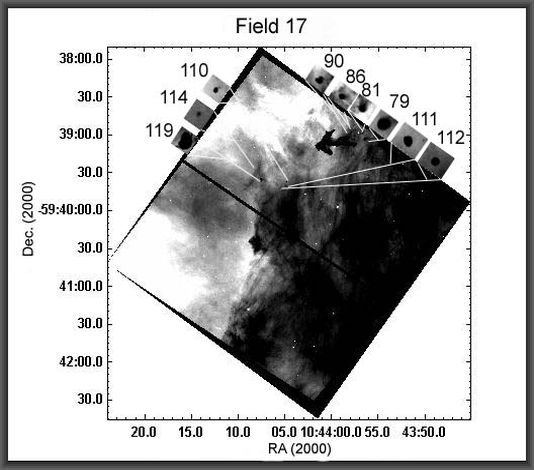}}
\resizebox{7cm}{!}{\includegraphics[angle=00]{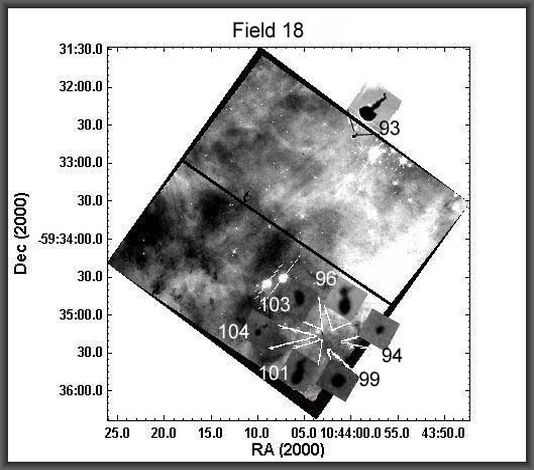}}
\resizebox{7cm}{!}{\includegraphics[angle=00]{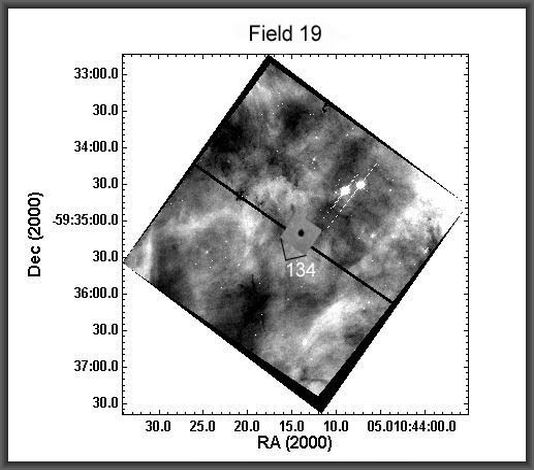}}
\resizebox{7cm}{!}{\includegraphics[angle=00]{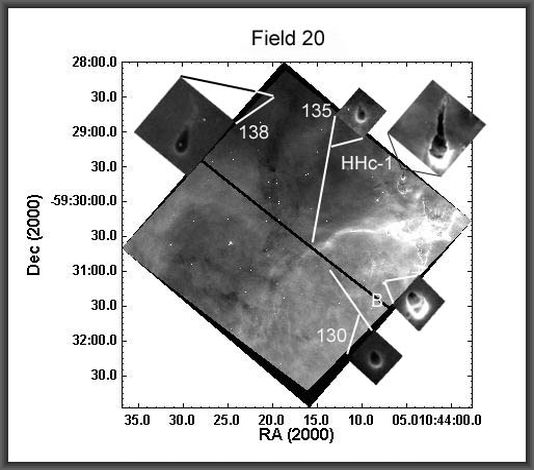}}
\resizebox{7cm}{!}{\includegraphics[angle=00]{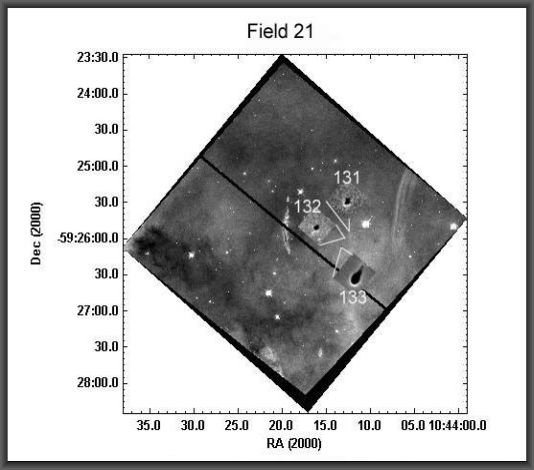}}
\resizebox{7cm}{!}{\includegraphics[angle=00]{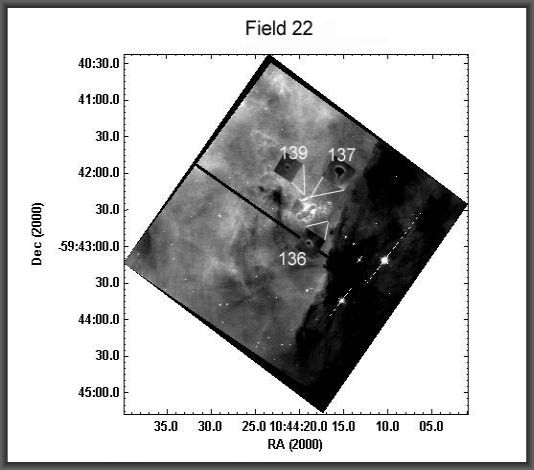}}
\resizebox{7cm}{!}{\includegraphics[angle=00]{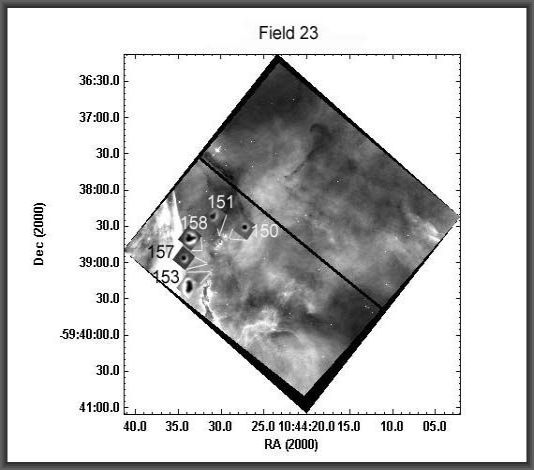}}
\resizebox{7cm}{!}{\includegraphics[angle=00]{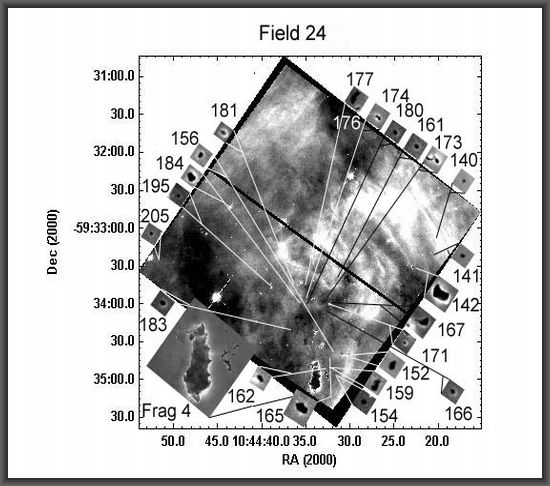}}
\caption{Fields 17 to 24}.
\label{fields3}
\end{figure*} 

\begin{figure*}[t] 
\centering
\resizebox{7cm}{!}{\includegraphics[angle=00]{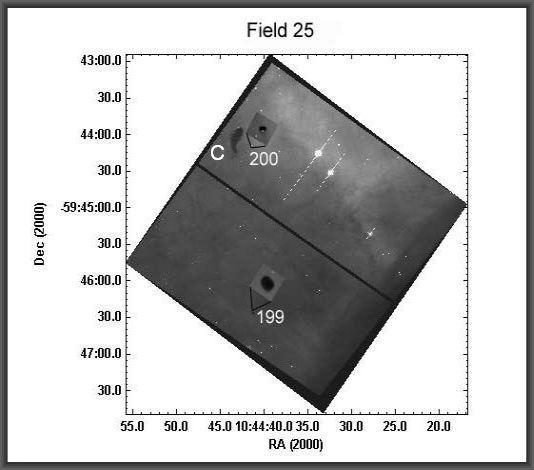}}
\resizebox{7cm}{!}{\includegraphics[angle=00]{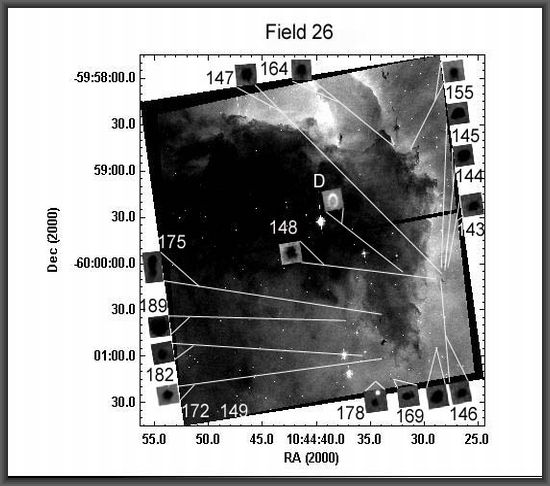}}
\resizebox{7cm}{!}{\includegraphics[angle=00]{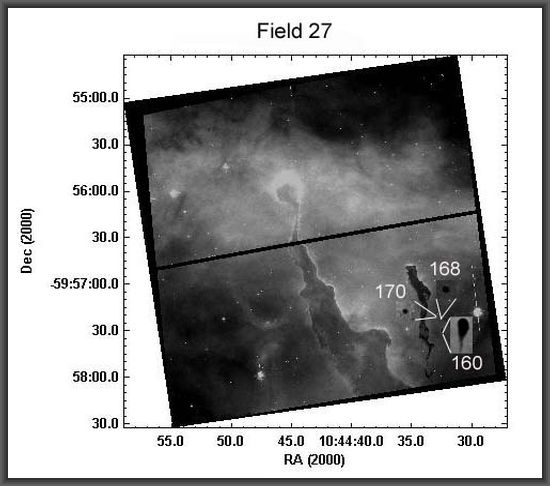}}
\resizebox{7cm}{!}{\includegraphics[angle=00]{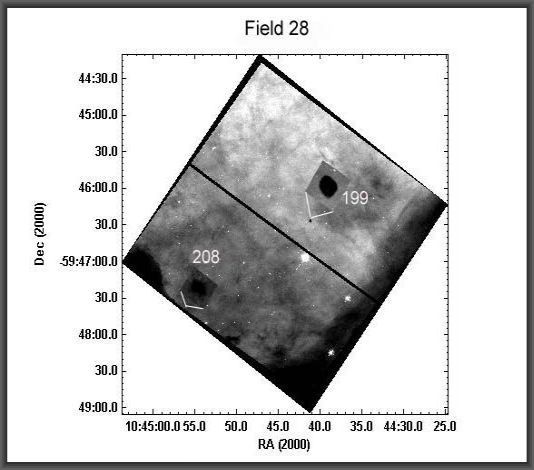}}
\resizebox{7cm}{!}{\includegraphics[angle=00]{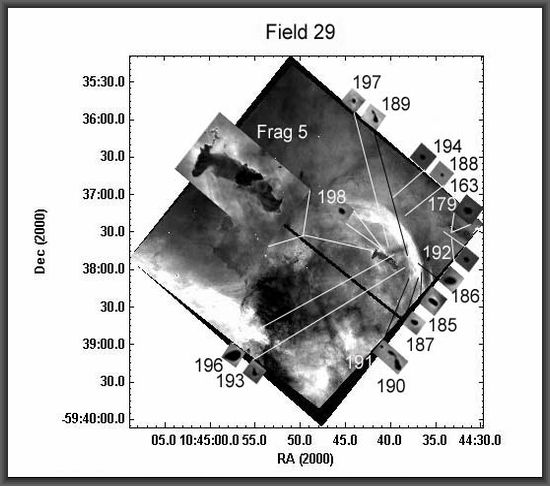}}
\resizebox{7cm}{!}{\includegraphics[angle=00]{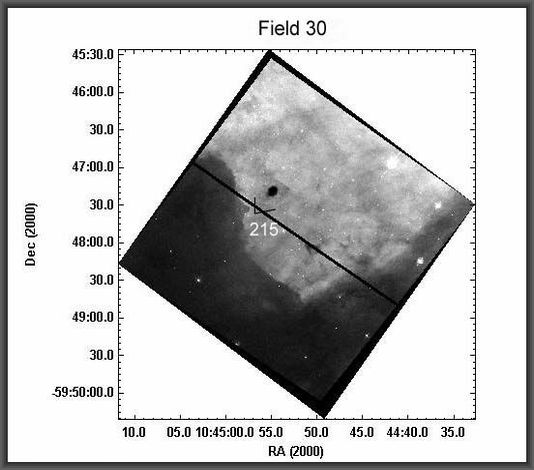}}
\resizebox{7cm}{!}{\includegraphics[angle=00]{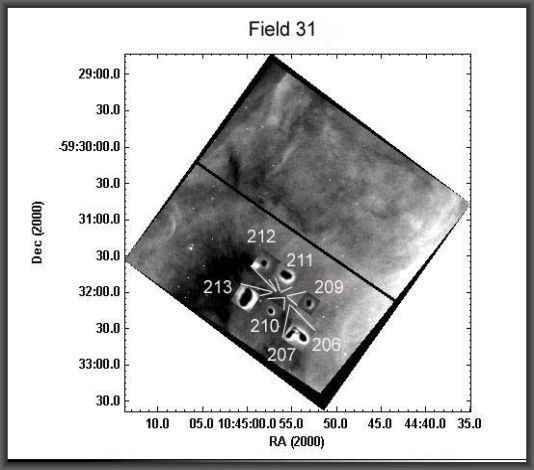}}
\resizebox{7cm}{!}{\includegraphics[angle=00]{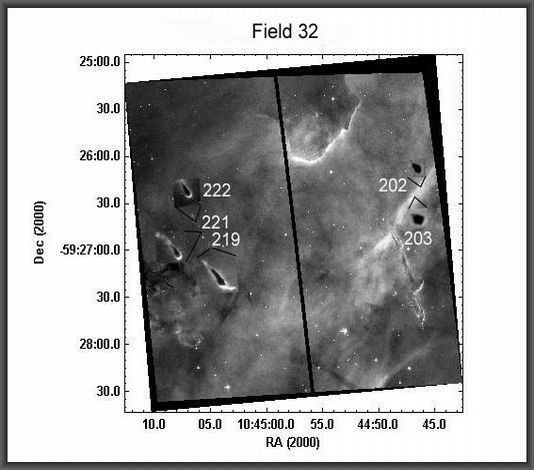}}
\caption{Fields 25 to 32.}
\label{fields4}
\end{figure*} 

\begin{figure*}[t] 
\centering
\resizebox{6.9cm}{!}{\includegraphics[angle=00]{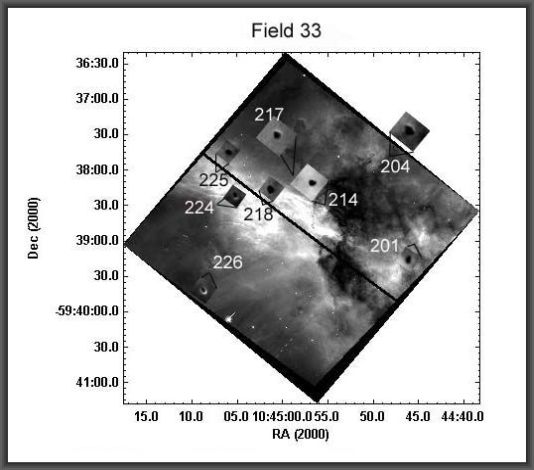}}
\resizebox{6.9cm}{!}{\includegraphics[angle=00]{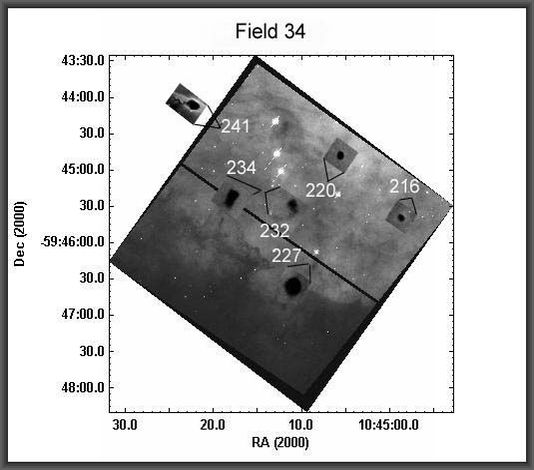}}
\resizebox{6.9cm}{!}{\includegraphics[angle=00]{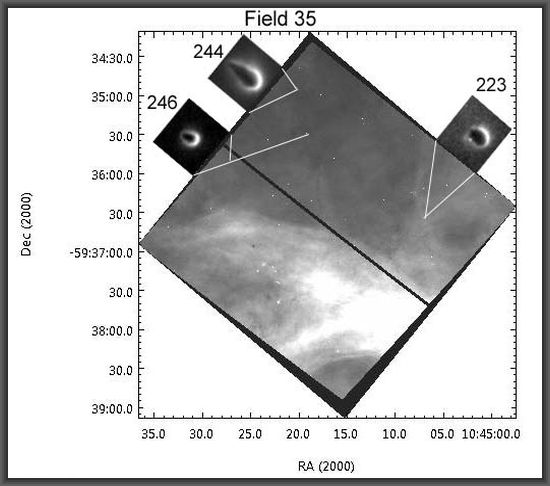}}
\resizebox{6.9cm}{!}{\includegraphics[angle=00]{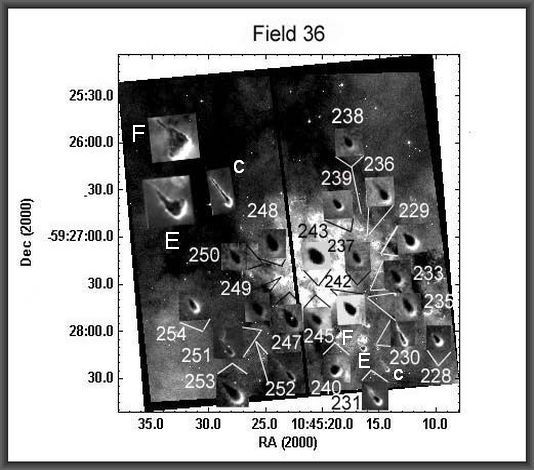}}
\resizebox{6.9cm}{!}{\includegraphics[angle=00]{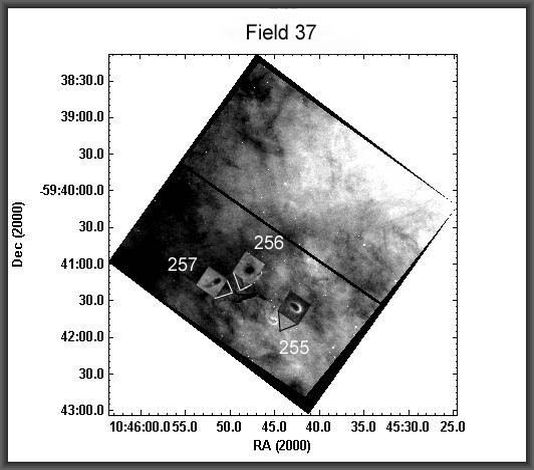}}
\resizebox{6.9cm}{!}{\includegraphics[angle=00]{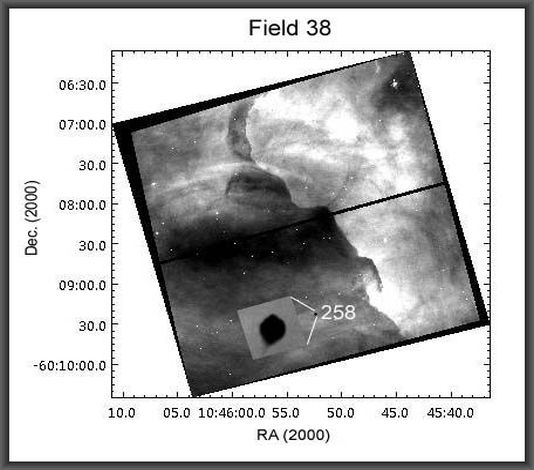}}
\resizebox{6.9cm}{!}{\includegraphics[angle=00]{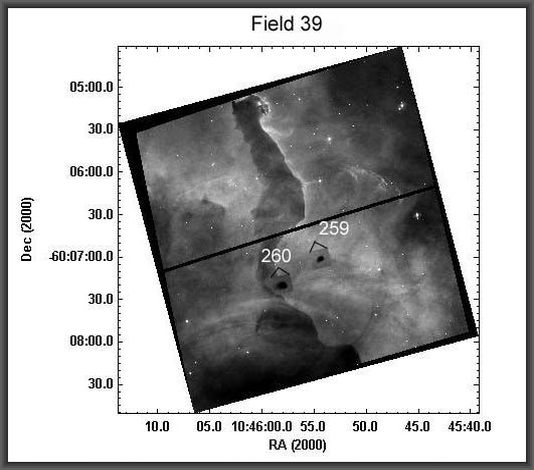}}
\resizebox{6.9cm}{!}{\includegraphics[angle=00]{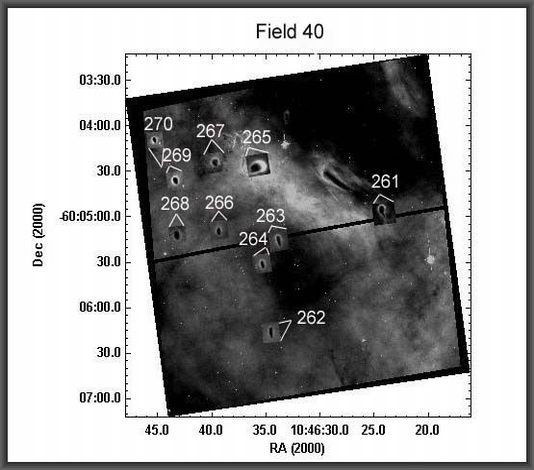}}
\caption{Fields 33 to 40.}
\label{fields5}
\end{figure*} 

\begin{figure*}[t] 
\centering

\resizebox{7cm}{!}{\includegraphics[angle=00]{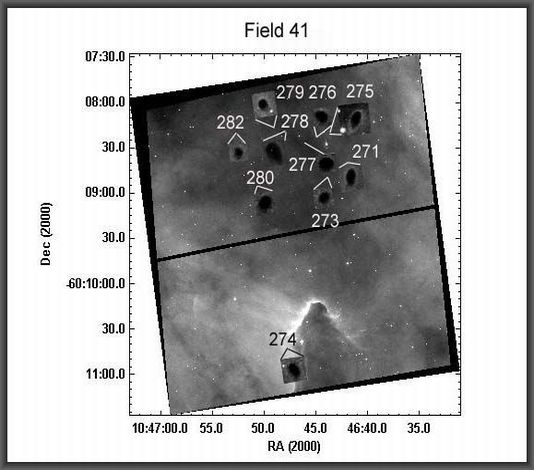}}
\resizebox{7cm}{!}{\includegraphics[angle=00]{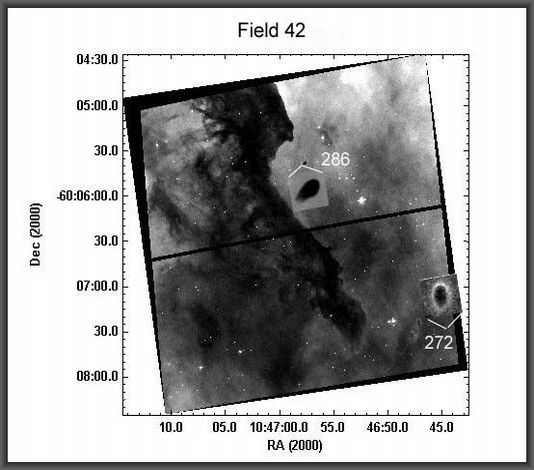}}
\resizebox{7cm}{!}{\includegraphics[angle=00]{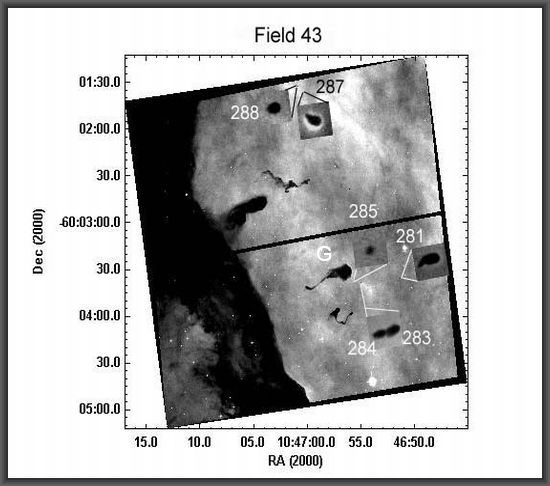}}
\caption{Fields 41 to 43.}
\label{fields6}
\end{figure*} 

\end{appendix}

\end{document}